\documentclass[useAMS,usenatbib,usegraphicx]{mn2e}
\usepackage{amsmath,mathtools,amsfonts,amssymb,float}
\usepackage[usenames,dvipsnames]{xcolor}
\usepackage{pdfpages}
\usepackage{soul}

\def\Msun{\hbox{$\rm\, M_{\odot}$}}
\def\vobs{\hbox{$\widetilde{\mathbf{v}}$}}
\def\eps{\hbox{$\epsilon$}}

\title [Quenching and galactic conformity] {A general approach to quenching and galactic conformity}
\author[L. Sin, S. Lilly, and B. Henriques]{Larry P. T. Sin$^{1}$\thanks{E-mail: sinp@phys.ethz.ch},
Simon J. Lilly$^{1}$, Bruno M. B. Henriques$^{1}$\vspace{0.4cm}\\
  {}$^{1}$Department of Physics, ETH Z\"urich, CH-8093 Z\"urich, Switzerland\\}
 
\begin{document}

\date{Submitted to MNRAS}

\volume{000}\pagerange{0000--0000} \pubyear{2019}

\maketitle

\label{firstpage}

\begin{abstract}
We develop a conceptual framework and methodology to study the drivers of the quenching of galaxies, including the drivers of galactic conformity. The framework is centred on the statistic $\Delta$, which is defined as the difference between the observed star-formation state of a galaxy, and a prediction of its state based on an empirical model of quenching. In particular, this work uses the average quenching effects of stellar mass $\mathrm{M_*}$ and local density $\delta$ to construct an empirical model of quenching. $\Delta$ is therefore a residual which reflects the effects of drivers of quenching not captured by $\mathrm{M_*}$ and $\delta$, or so-called \lq hidden variables\rq. Through a toy model, we explore how the statistical properties of $\Delta$ can be used to learn about the internal and external hidden variables which control the quenching of a sample of galaxies. We then apply this analysis to a sample of local galaxies and find that, after accounting for the average quenching effects of $\mathrm{M_*}$ and $\delta$, $\Delta$ remains correlated out to separations of 3 Mpc. Furthermore, we find that external hidden variables remain important for driving the residual quenching of low-mass galaxies, while the residual quenching of high-mass galaxies is driven mostly by internal properties. These results, along with a similar analysis of a semi-analytical mock catalogue, suggest that it is necessary to consider halo-related properties as candidates for hidden variables. A preliminary halo-based analysis indicates that much of the correlation of $\Delta$ can be attributed to the physics associated with individual haloes.
\end{abstract}


\section{Introduction}
\label{sec:intro}

A striking characteristic of galaxies in the Universe is the bimodality in the distribution of their star-formation rates \citep{Kauffmann2003, Baldry2004, Brinchmann2004}. On one hand, there is a population of galaxies which are actively forming stars. These galaxies have star-formation rates which are strongly correlated with their stellar mass, producing a so-called \lq Main Sequence\rq\ \citep{Brinchmann2004, Daddi2007, Elbaz2007, Noeske2007, Salim2007}. On the other hand, there is also a population of passive galaxies, for which the rate of star-formation is suppressed by one to two orders of magnitude relative to the Main Sequence.

The existence of this bimodality indicates that, at the end of the active portion of their existence, galaxies cease their star-formation almost completely. Understanding the transition from the star-forming state to the passive state, a transition known as \lq quenching\rq, remains a major goal in the field of galaxy evolution. It is clear that both the stellar mass of a galaxy, and the environment in which it lives, play key roles in quenching, in the sense that the fraction of quenched galaxies increases as a function of both of these variables \citep{Peng2010}. Additionally, it is clear that the relation between galaxies and their host dark matter halo plays an important role in quenching. The correlation between mass and quenching is thought to indicate that central galaxies, which formed and evolved at the centre of their halo, experience the disruption of their reservoirs of star-forming gas via energetic feedback from their Active Galactic Nucleus \citep[hereafter AGN;][]{Baldry2006, Bower2006, Croton2006, Bluck2014, Henriques2017}. On the other hand, the correlation between environment and quenching is primarily attributed to the quenching of satellite galaxies, which orbit within the halo of another, more massive galaxy. Satellite galaxies are thought to experience disruptions to their supply of star-forming gas due to physical effects associated with the dense halo environment, such as ram-pressure stripping, tidal
stripping, and strangulation \citep{Gunn1976, Larson1980, Moore1996, Abadi1999}.

There are also indications that the quenching of centrals and the quenching of satellites may be closely linked: \cite{Weinmann2006} found that, for groups of galaxies at a fixed halo mass, there exists a correlation between the star-formation state of centrals, and that of their satellites. That is, passive centrals tend to have passive satellites, and star-forming centrals tend to have star-forming satellites. The existence of this correlation, which they named \lq galactic conformity\rq, could indicate that the quenching of satellites is to some extent causally dependent on the quenching of their centrals, or vice versa. Alternatively, it could be an indication that the quenching of galaxies within groups, central and satellites alike, is influenced by a group-wide mechanism. For instance, it has been proposed that a sufficiently energetic AGN can influence the star-formation of an entire group by heating the intergalactic gas \citep{Weinmann2006, Kauffmann2013}. 

Generally, conformity can be thought of as an indication of the presence of a group-wide \lq hidden common variable\rq, which either facilitates the causal dependence of central- and satellite-quenching, or which influences the quenching of an entire group. A hidden variable which is correlated on the scale of individual haloes, but varies from halo to halo, could explain the existence of conformity.

It then follows that a sample of groups which are selected to be \lq matched\rq\ in the relevant hidden variables should exhibit no remaining conformity, in the sense that satellites of quenched centrals should not be preferentially quenched relative to satellites of star-forming centrals in the matched sample. Following this reasoning, \cite{Knobel2015} evaluated candidates for hidden variables by evaluating the conformity signal in SDSS groups at matched halo mass, and subsequently matching additional variables, namely the normalized group-centric distance, the local density, the stellar mass of the central, and the stellar mass of the satellite. Even for a sample of haloes matched in these five variables, there remained a significant conformity signal.

A surprising development was the work of \cite{Kauffmann2013}, who presented observational evidence for a strong conformity signal extending out to very large distances. However, the works of \cite{Sin2017} and \cite{Tinker2018} illustrated that methodological choices in the analysis of conformity made by \cite{Kauffmann2013} can artificially produce a misleadingly strong, long-range conformity signal. In particular, \cite{Sin2017} identified three key methodological features which strongly amplified the correlation between the specific star-formation rates of satellite galaxies of very massive haloes. After accounting for this, the evidence for the existence of conformity beyond individual haloes became statistically questionable.

The methodological points raised in \cite{Sin2017} also demonstrated the difficulty in interpreting the physical meaning of a measurement of conformity. In particular, it was found that the interpretation of the spatial scale of a conformity signal is not straightforward: given that a generic sample of haloes will contain haloes with a range of sizes, the physical meaning of the scale of a conformity signal should be considered with regard to the range of environments from which the signal originates. Similarly, given that the statistics of the star-formation states of a galaxy sample will depend on the distribution of the physical properties of the galaxies, the magnitude of a measurement of conformity (e.g. differences in quenched fractions) should only be interpreted with regard to the distribution of galaxy properties within the sample.

These difficulties motivate a careful consideration of how conformity should be defined, how it should be measured, and how the results should be interpreted. In this work, we present a framework of analysis which allows us to measure and meaningfully interpret measurements of conformity by explicitly accounting for the dependence of star-formation states on known drivers of quenching, and by framing conformity in more general terms as the result of the spatial correlation of hidden variables.

This paper is organized as follows: In Section \ref{sec:concept}, we describe our conceptual framework regarding conformity, and thereby motivate the development a new method of analysis. In Section \ref{sec:data}, we describe the observational and simulated data used in this work. In Section \ref{sec:meth}, we describe our analytical framework, and present tests of its ability to recover conformity in simulated data. In Section \ref{sec:results}, we present the results from our analysis of observational and simulated data. In Section \ref{sec:discussion}, we discuss the implications of our results on the nature of conformity. Finally, in Section \ref{sec:summary}, we summarize our conclusions.

We use a $\Lambda\mathrm{CDM}$ cosmology with $\Omega_{\Lambda}= 0.7$, $\Omega_\mathrm{M} = 0.3$, and $H_0 = 67.7\;\mathrm{km\,s}^{-1}\,\mathrm{Mpc}^{-1}$. The logarithms in this work are in base 10, and we use the dimensionless unit \lq dex\rq\ to denote the anti-logarithm. That is to say, a multiplicative difference by a factor of $10^n$ in linear space is equal to an additive difference of $n$ dex in logarithmic space.

\section{Conceptual framework}
\label{sec:concept}

\subsection{A review of \citet{Knobel2015}}
\label{subsec:review_knobel}

One key difficulty in the study of conformity is in understanding the significance of the correlations between the star-formation state of a central galaxy, and that of its satellites. Consider the example of a sample of galaxy groups with a range of halo masses: in general, a correlation between the star-formation states of centrals and those of their satellites will exist, simply due to the fact that galaxies in low-mass haloes (centrals and satellites alike) are in general less likely to be quenched, while those in high-mass haloes are in general more likely to be quenched. The significance of conformity, as measured by \cite{Weinmann2006}, is that the correlation between the star formation of centrals and satellites persists even when the halo masses of groups are controlled for. That is, the existence of conformity indicates that the quenching of galaxies within haloes is not well-described by halo mass alone. In general, the detection of a significant conformity signal will require a sample of haloes which is controlled for in all other variables which are relevant to quenching (e.g. stellar mass, environment). 

\cite{Knobel2015} developed a framework to do so via the \eps\ statistic. For a sample of galaxies, the statistic is defined as \begin{equation} \label{eq:epsilon}
\eps\ \coloneqq \frac{\hat{f}_q-f_{q,\mathrm{model}}}{1-f_{q,\mathrm{model}}},
\end{equation} where $\hat{f}_q$ is the observed fraction of quenched galaxies in the sample, and $f_{q,\mathrm{model}}$ is what one would expect $f_{q}$ to be based on the modelling of the quenching effects of various known galaxy properties. To draw from the previous example: as it is known that the host halo mass of a galaxy, denoted as $\mathrm{M}_{h}$, is related to its likelihood of being quenched, one may wish to account for this before considering the significance of the correlation of star-formation states. A simple approach to do this would be to first constrain the average quenching effect of $\mathrm{M}_{h}$ by measuring, for a sample of galaxies, the quenched fraction at every point in $\mathrm{M}_{h}$-space. Given this measurement, $f_q(\mathrm{M}_{h})$, one can predict the quenched fraction of a subsample based on its distribution of halo masses, and use this as $f_{q,\mathrm{model}}$ in equation (\ref{eq:epsilon}). Naturally, this will not be a perfect prediction, as there are other variables beside $\mathrm{M}_{h}$ which may drive quenching. \eps\ quantifies the difference between the observed and the predicted quenched fraction, and reflects the quenching effects of variables which are \lq orthogonal\rq\ to that of $\mathrm{M}_{h}$. We refer to these variables as \lq hidden variables\rq\ of quenching. 

When considering a sample of galaxy groups, conformity can be quantified by the difference in \eps\ between the satellites of passive centrals and those of star-forming centrals, as was done in \cite{Knobel2015}. One-halo conformity can then be thought of as an indication that there are hidden variables which are correlated on halo scales, and which affect the quenching of both centrals and satellites. \cite{Knobel2015} referred to these as \lq hidden common variables\rq.

There are two aspects of this framework which can be developed further. The first is to explore and clarify the meaning of the statistic $\hat{f_q}-f_q$. Qualitatively, it is understood to be the residual of the data relative to the model. However, quantitatively, it is unclear what a particular value of $\hat{f_q}-f_q$ indicates about a sample of galaxies, or its associated hidden variables. Therefore, we will first explicate the meaning of this statistic. The second aspect to develop is that, in \cite{Knobel2015}, and indeed in many other works which deal with conformity, one is restricted to thinking in terms of haloes. This presents a difficulty in thinking about conformity on larger scales, where one wishes to consider the correlation of \eps\ between galaxies residing in different haloes (sometimes referred to as \lq two-halo\rq\ conformity). Therefore, the second development which we will present is a more general framework, within which one can meaningfully quantify both the strength and the spatial scale of conformity.

\subsection{The meaning of $\hat{f_q}-f_q$}
\label{subsec:fq_meaning}

The quenched fraction $f_q$ is often referred to as a quenching probability, and correspondingly, $\hat{f_q}-f_q$ is often referred to as an excess quenching probability, or more commonly a relative quenching efficiency \citep{Peng2010, Omand2014, Knobel2015}. The meaning of these terms are unclear; in fact, the concept of \lq probability\rq\ is at odds with the presumably deterministic nature of quenching. 

In reality, the specific star-formation rate ($\mathrm{SFR/M_*}$; hereafter SSFR) of a galaxy is presumably completely determined by the physical properties and the physical processes associated with that galaxy. These are thought to be diverse in nature, from internal properties such as molecular gas mass and AGN activity \citep{Kennicutt1998,Croton2006,Henriques2017}, to larger-scale phenomena typically associated with the physics of galaxy clusters \citep{Gunn1972,Moore1996,Abadi1999}; hereafter, we will simply refer to these properties and processes as \lq variables\rq, since each of them can typically be characterized by a single number (e.g. $\mathrm{M_{H_2}}$, $\lambda_{\mathrm{Edd.}}$, $t_{\mathrm{infall}}$). If one had a complete understanding of galaxy evolution, and could observe all of the relevant variables associated with a galaxy, one should in principle be able to predict its SSFR without any uncertainty. To frame this more simply, one could define the star-formation state of a galaxy (i.e. star-forming or quenched) based on a threshold in SSFR, and by the same logic, one should in principle be able to predict exactly whether a galaxy is star-forming or quenched. Symbolically, this can be expressed as \begin{equation} \label{eq:quenching}
q=g(\mathbf{v});\ q\in \lbrace 0,1\rbrace,
\end{equation} where $\mathbf{v}$ is the collection of variables which are relevant to quenching, $g$ represents the (probably very complicated) physical relation which maps these variables to the galaxy's star-formation state, and 0 and 1 respectively represent star-forming and quenched\footnote{The usage of 0 and 1 to represent star-forming and quenched is a matter of consistency. When defined in this way, the mean $q$ of a sample is identical to its $\hat{f}_q$.}. We assume that $q$ can be observed without error. When expressed in these terms, the difficulties in our attempt to understand quenching (i.e. the behaviour of $q$) can be expressed as follows: our understanding of galaxy evolution is incomplete (we do not know $g$), and we do not observe all of the relevant variables (we do not know $\mathbf{v}$).

However, we do have some knowledge of observable variables which are relevant to quenching, which we denote as \vobs. A simple way to express the dependence of $q$ on these variables would be to estimate, at every point in {\vobs}-space, the observed fraction of quenched galaxies at that point\footnote{While conceptually straightforward, this is in practice not a trivial task. See Section \ref{subsec:fq} for details on how this might be done in practice.}. Conversely, for a galaxy with a given \vobs, its $f_q(\vobs)$ can be thought of as a non-parametric prediction of what the star-formation state of that galaxy might be, based on our current state of knowledge. That is to say that for a galaxy with a given \vobs, while its $q$ is in reality determined by some unknown dependence on $\mathbf{v}$, we can only make an approximate prediction of what its $q$ might be based on its measured observables, \vobs, and the general dependence on these variables, $f_q(\vobs)$.

An interesting quantity to then consider is the difference between the actual (and presumably accurately measured) star-formation state of a galaxy, and our best prediction of its state, \begin{equation}
\Delta \coloneqq q-f_q.
\end{equation} Given that quenching is deterministic (i.e. there is no element of randomness involved), $\Delta$ then exclusively reflects the presence of quenching effects of variables which we have not accounted for via $f_q(\vobs)$. That is, $\Delta$ quantifies the degree of our ignorance of quenching. This is the central point of this framework: $\Delta$ reflects our inability to empirically predict the star-formation state of a galaxy, which arises either from a lack of knowledge of the relevant properties, or a lack of knowledge of how these properties affect the evolution of a galaxy.

We would like to understand the meaning of the statistical properties of $\Delta$. By construction, $f_q$ at any point in {\vobs}-space is the mean $q$ at that point. From this property\footnote{The distribution of $q$ at a given \vobs\ can be described by a Bernoulli distribution, with the parameter $p$ being equivalent to $f_q(\vobs)$. The following statements about the behaviour of $\Delta$ follow from the properties of this distribution.}, it follows that $\Delta$ is on average 0 at any point in {\vobs}-space, and consequently has a sample average of 0. A more interesting quantity to consider is the average magnitude of the deviation of $q$ from $f_q$, i.e. the variance of $\Delta$. Again, it follows from the construction of $f_q$ that at a given point in {\vobs}-space, the variance is $f_q(1-f_q)$. Therefore, the variance of $\Delta$ of a galaxy sample is given by taking the average over the \vobs-distribution of the sample, \begin{equation} \label{eq:sigam_delta}
\sigma_{\Delta}^2=\langle f_q(1-f_q)\rangle.
\end{equation} 

For the sake of illustration, consider one limiting case in which we know nothing about the evolution of galaxies, and are therefore unable to model $f_q$ as a function of any variables. In this case, our best prediction of the state of any galaxy would simply be the sample mean $q$, denoted as $\bar{q}$, and so we set the $f_q$ of all galaxies in the sample to be $\bar{q}$. It then follows that the variance is given by $\sigma_{\Delta}^2=\bar{q}\,(1-\bar{q})$.

Consider now the other limiting case in which we understand the evolution of galaxies completely. In this case, we are able to set $g(\mathbf{v})$ as our $f_q$. Since the state of each galaxy is predicted perfectly, $q$ never deviates from $f_{q}$, so $\Delta$ is identically 0, and $\sigma_{\Delta}^2=0$.

Generally, for an $f_q(\vobs)$ which is only an approximate description of quenching, we are bound by these two cases. Note the consistency of the formula for $\sigma_{\Delta}^2$ with the intuition that we have developed; as our understanding of galaxy evolution becomes more complete, we should have greater confidence in predicting whether a galaxy should be quenched or not. As a result, the distribution of $f_q$ in the sample tends to move away from $\bar{q}$, and closer to either 0 or 1, and so $\langle f_q(1-f_q)\rangle$ tends to 0. 

In Section \ref{subsec:toy_model}, we will develop our understanding of the behaviour of $\sigma_{\Delta}^2$ further by exploring a toy model of quenching. Then, in Section \ref{subsec:concept_remarks}, we remark upon some important subtleties and limitations of this framework of analysis.

\subsection{The meaning of conformity}
\label{subsec:conformity_meaning}

Recall that in the \cite{Knobel2015} framework, conformity is measured in terms of the difference in \eps\ between the satellites of passive centrals and those of star-forming centrals; it is thought of as an indication that there are hidden variables which are correlated on halo scales, and which affect the quenching of both centrals and satellites. In principle, the same measurement can be made using $\Delta$ instead of \eps. However, such an approach is intrinsically halo-based, and one would therefore only learn about the presence of hidden variables which are correlated on the scales of individual haloes. This presents a difficulty in thinking about conformity beyond the scale of individual haloes, where one wishes to consider the correlation of $\Delta$ between galaxies residing in different haloes (i.e. \lq two-halo\rq\ conformity).

The generalization which we present here stems from the fact that hidden variables do not necessarily have to be thought of as properties of haloes, but can rather be thought of as a property of the space in which galaxies reside. That is, hidden variables can instead be characterized by a spatially correlated field. Note that \lq field\rq\ here is not meant in any physical sense, but strictly in the mathematical sense that it has a value at every point in space. For our purposes, the term can be used somewhat more loosely: the hidden variable field only needs to be defined at every point where there is a galaxy. The reasoning then follows as before, that quenching is driven by hidden variables, and $\Delta$ will therefore exhibit a spatial correlation (i.e. conformity) which follows the spatial correlation of the hidden variables.

Conformity can thus be quantified via the spatial correlation function of $\Delta$. For a scalar field $f$ defined over spatial coordinates $\bf{x}$, the correlation function $\xi_{f}$ at separation $R$ is defined as the expectation value of the product of $f$ for two points separated by $R$. That is, \begin{equation} \label{eq:corrfun} 
\xi_{f}(R) \coloneqq E\left[f(\mathbf{x})f(\mathbf{x}\rq)\right],
\end{equation} where $\left|\bf{x-x\rq}\right|=R$. 

For a field with zero mean, the variance and the correlation function are related: $\xi_{f}(0) = E\left[f^2\right]=\sigma_f^2$. However, this relation holds even for a field which is completely uncorrelated (e.g. a field of noise). An analogous, but more relevant quantity to consider is the limiting value of $\xi_f(R)$ as $R$ approaches zero, \begin{equation} \label{eq:xi_0} 
\xi_{f,0} \coloneqq \lim_{R\to 0}\xi_f(R).
\end{equation} This quantity recovers $\xi_{f}(0)$ for a continuous (e.g. perfectly correlated) field, but is in general bound between $\sigma_f^2$ and zero, depending on the nature of the field's correlation.

As $\Delta$ is constructed to have zero mean, the above properties of the correlation of a general $f$ also hold for $\Delta$. In this case, $\xi_{\Delta}$ would be driven by hidden variables which are spatially correlated. Note that as $f_q$ becomes a better description of quenching, $\xi_{\Delta}$ behaves analogously to $\sigma_{\Delta}^2$: while $\sigma_{\Delta}^2$ decreases as our general description of quenching (i.e. $f_q$) improves, $\xi_{\Delta}$ decreases only when our description of spatially correlated variables improves.

There is an additional interesting statistical property of $\Delta$, especially for the case of $\Delta = q-\bar{q}$. Since the full set of $\mathbf{v}$ presumably consists of both spatially correlated (i.e. environmental) drivers, as well as uncorrelated (i.e. internal) ones, we expect $\xi_{\Delta,0}$ to only be a fraction of $\sigma_{\Delta}^2$. This ratio, which we denote as \begin{equation} \label{eq:coherence_factor}
\gamma_{\Delta}\coloneqq\frac{\xi_{\Delta,0}}{\sigma_{\Delta}^2},
\end{equation} reflects (in a non-trivial way) the relative importance of spatially correlated (i.e. environmental) variables and spatially uncorrelated (i.e. internal) variables to quenching. By the same reasoning, as $f_q$ becomes a more complete description of quenching by accounting for more variables, $\gamma_{\Delta}$ would reflect (again, in a non-trivial way) the importance of environmental effects relative to internal effects for the remaining hidden variables.

When conformity is thought of in this way, its significance becomes clear. First, the spatial scale of the correlation function $\xi_{\Delta}(R)$ reflects the physical scale of the environmental hidden variables which are driving quenching. Second, by comparing the strength of conformity (i.e. $\xi_{\Delta,0}$) relative to our overall ignorance (i.e. $\sigma_{\Delta}^2$), one can learn about the relative importance of environmental effects and internal effects to quenching.

In the following subsection, we explore the statistical properties of $\Delta$ further through a toy model.

\subsection{A toy model of quenching}
\label{subsec:toy_model}

\begin{figure}
\centering
\includegraphics[width=8.6cm]{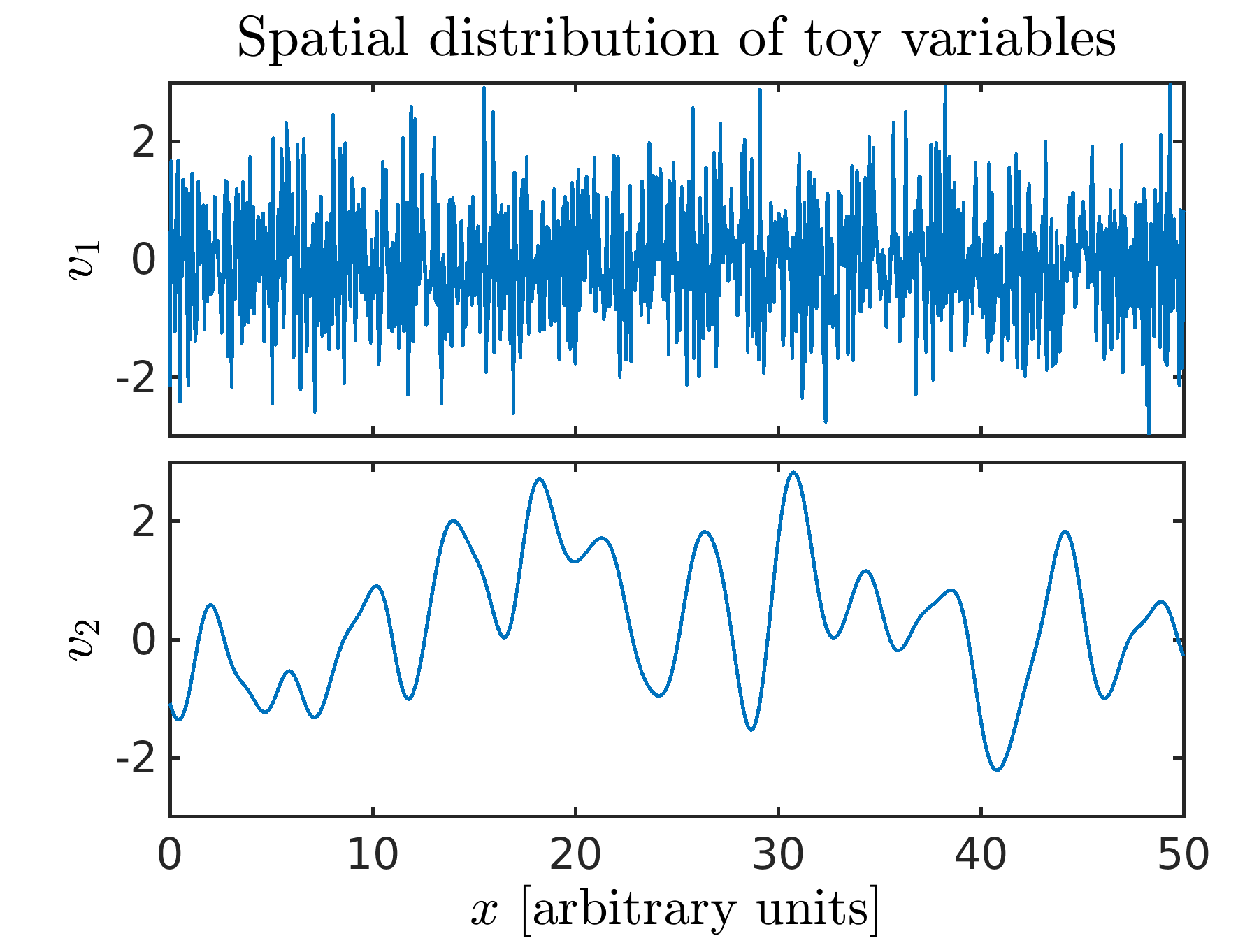}
\caption{Visualization of the spatial distribution of galactic variables in the toy model. $v_1$ is generated as Gaussian noise with $\mu_{v_1}=0$ and $\sigma_{v_1}^2=1$, while $v_2$ is generated as a Gaussian random field with $\mu_{v_2}=0$, $\sigma_{v_2}^2=1$, and $\xi_{v_2}(R)=\sigma_{v_2}^2\,\mathrm{exp}(-R^2/2)$.}
\label{fig:toy_model_vars}
\end{figure}

\begin{figure}
\centering
\includegraphics[width=8.6cm]{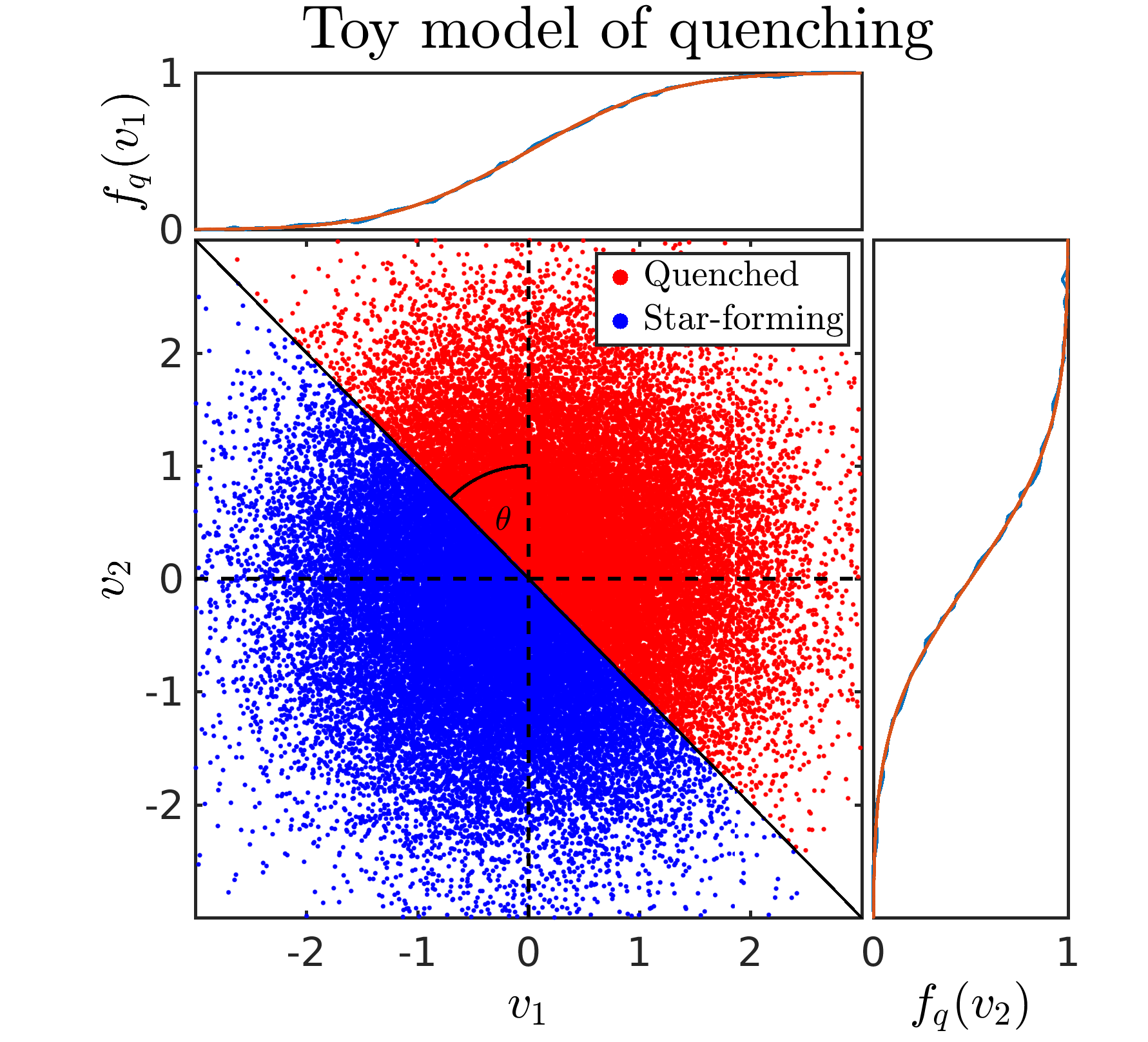}
\caption{Visualization of quenching within the toy model, for the case where the relative importance of the correlated and the uncorrelated variables are set to be equal by setting $\theta=\pi/4$. The main panel shows the mapping between the galaxies' variables and their star-formation states. In the marginal panels, the smooth orange line shows the analytical form of $f_q$ as a function of one variable, while the blue line shows the measured $f_q$.}
\label{fig:toy_model_quench}
\end{figure}

\begin{figure*}
\centering
\includegraphics[width=17.9cm]{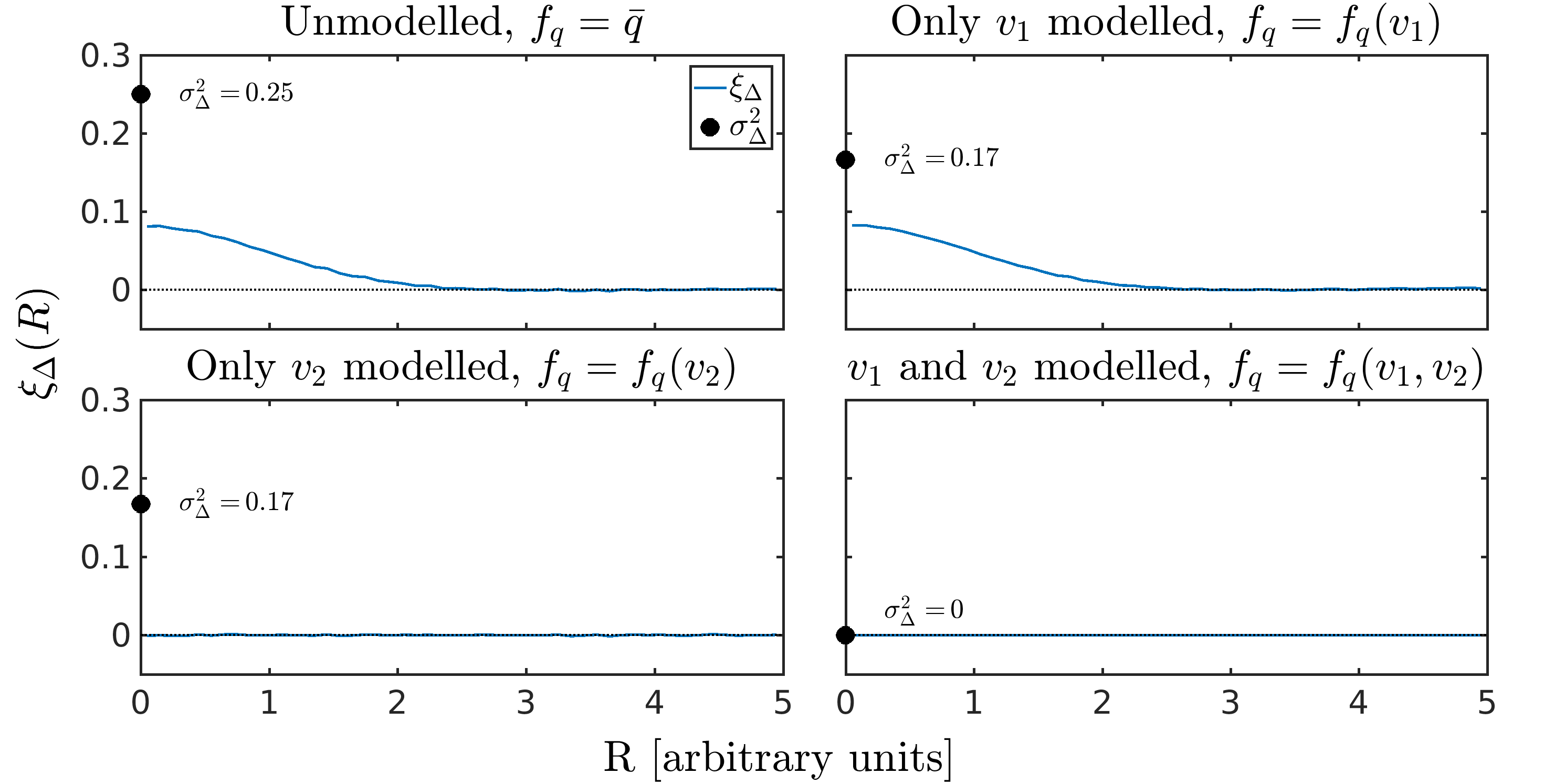}
\caption{The measured correlation functions of $\Delta$ resulting from the toy model, where the panels make use of different $f_q$'s to reflect different states of knowledge. The top-left panel reflects the state of complete ignorance, where neither $v_1$ nor $v_2$ are known, and so our best prediction of the star-formation state for a given galaxy, $f_q$, is simply the sample average, $\bar{q}$.  The top-right and the bottom-left panels both reflect a case in which only one variable is known, and so the corresponding $f_q(v)$ is used as the prediction. Finally, the bottom-right panel reflects the state of complete knowledge, in which both $v_1$ and $v_2$ are known, and so the best prediction is identical to the actual mapping of $\mathbf{v}$ to $q$. In each panel, the correlation function of its corresponding $\Delta$ is plotted in blue, while the variance of $\Delta$ is indicated by the black point. In each case, the correlation function reflects the effects of the remaining hidden variable(s) which are spatially correlated, while the variance reflects the overall effects of the remaining hidden variable(s).}
\label{fig:toy_model_cf}
\end{figure*}

\begin{figure}
\centering
\includegraphics[width=8.6cm]{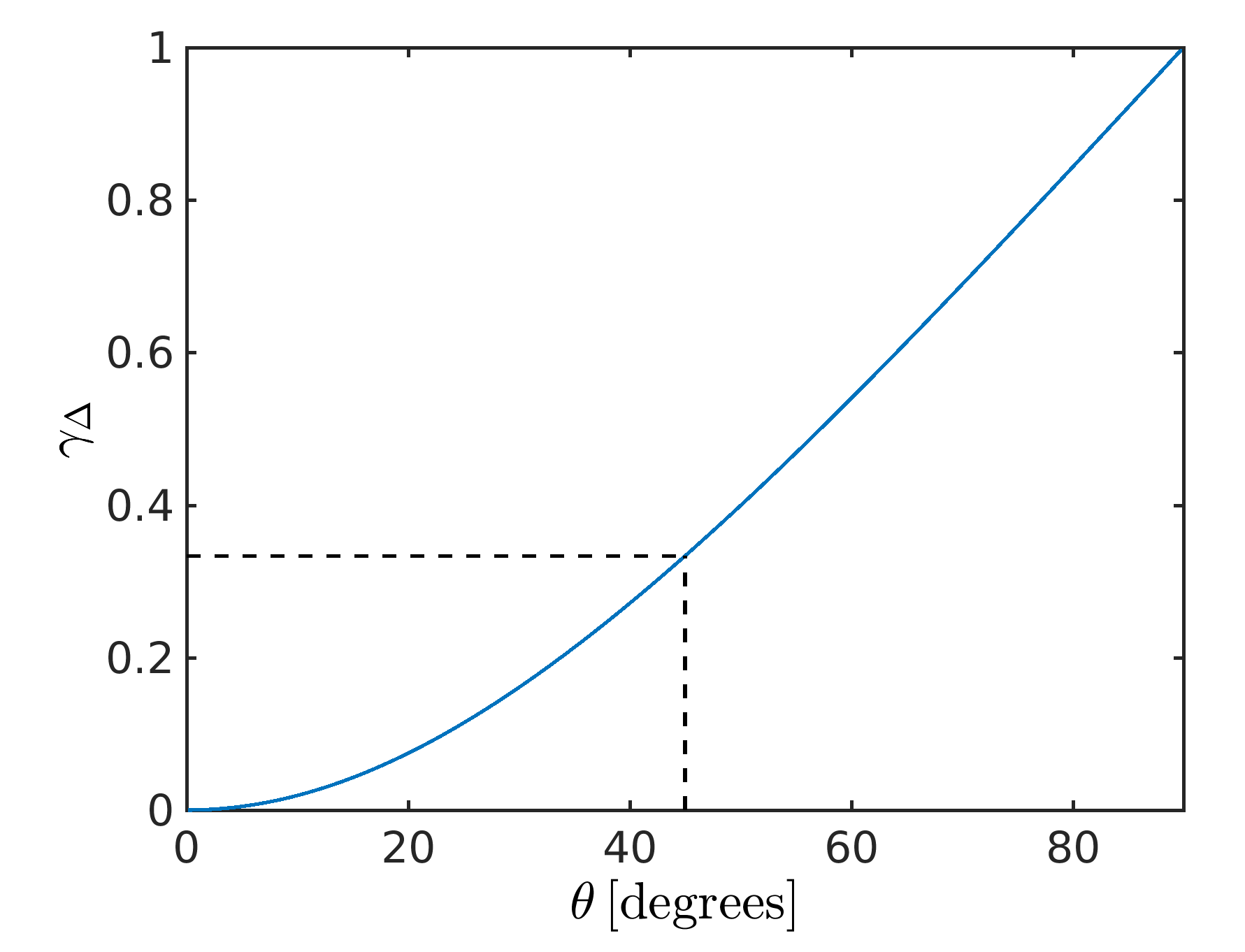}
\caption{The graph of the analytical relation between $\gamma_{\Delta}$ and $\theta$ for the case of $\Delta=q-\bar{q}$. The dashed black lines indicate the configuration used in the toy model, where the spatially correlated and the uncorrelated variables are given equal importance to quenching.}
\label{fig:toy_model_gamma}
\end{figure}

In this subsection, we attempt to help the reader develop an intuitive understanding of our framework by considering a heuristic \lq toy\rq\ model of quenching.

Consider a toy universe in which galaxies are distributed uniformly in a 1-dimensional space, with position denoted by $x$, and have only two variables which are relevant to quenching, denoted by $\mathbf{v}=\lbrace v_1,v_2\rbrace$. We define $v_1$ to be a Gaussian random field with $\mu_{v_1}=0$, $\sigma_{v_1}^2=1$, and no autocorrelation. $v_1$ is intended to represent a driver of quenching which is internal to a galaxy, and therefore has no spatial correlation. On the other hand, we define $v_2$ to be a Gaussian random field with $\mu_{v_2}=0$, $\sigma_{v_2}^2=1$, and autocorrelation $\xi_{v_2}(R)=\sigma_{v_2}^2 \,\mathrm{exp}(-R^2/2)$. $v_2$ is intended to represent an environmental or external driver of quenching, and hence it is spatially correlated. A visualization of the spatial distribution of these variables is presented in Figure \ref{fig:toy_model_vars}.

Within this toy model, we set the star-formation state of galaxies to be determined by $v_1$ and $v_2$ in the following way: \begin{equation}
q=H\left[\frac{v_1}{\mathrm{tan}(\theta)}+v_2\right],
\end{equation} where $H$ is the Heaviside step function, and $\theta$ is a parameter which determines the relative importance of $v_1$ and $v_2$ to quenching. This model is represented in Figure \ref{fig:toy_model_quench}. The main panel of the Figure shows the joint distribution of the variables, while the colours reflect the mapping of these variables to star-formation states. In particular, this Figure reflects the case in which the relative importance of $v_1$ and $v_2$ are set to be equal by setting $\theta = \pi/4$, and equation (\ref{eq:quenching}) simply reduces to $q=H\left[v_1+v_2\right]$. The panels on the margins show the apparent fraction of quenched galaxies as a function of one variable, when the other variable is marginalized out. These $f_q$'s are analogous to the $f_q(\vobs)$ discussed in Section \ref{subsec:fq_meaning}; that is, they are the \lq best guess\rq\ of a galaxy's state in the case where there is only partial knowledge of the full set of $\mathbf{v}$. Notice that under these states of partial knowledge, the idea of quenched fractions emerges naturally, not as a result of \lq probability\rq\ or \lq relative efficiency\rq, but as a result of ignorance.

Given this setup, we then explore the behaviour of the statistics of interest, $\sigma_{\Delta}^2$ and $\xi_{\Delta,0}$, in Figure \ref{fig:toy_model_cf}. In each panel, a different state of knowledge is assumed, and the corresponding $f_q$ is used in the evaluation of $\Delta$.

The top-left panel reflects the state of complete ignorance, where neither $v_1$ nor $v_2$ are known. In this case, our best prediction of the star-formation state for any given galaxy is simply the sample average, $\bar{q}$. The correlation function of the corresponding $\Delta$ is plotted in light blue, while the value of the variance is indicated by the black point. As a result of the spatial correlation of $v_2$, $\Delta$ also has spatial correlation. However, due to the presence of the uncorrelated variable, $v_1$, the correlated variable only contributes to part of the total variance, and $\xi_{\Delta,0}<\sigma_{\Delta}^2$.

The top-right panel reflects the case in which only $v_1$ is known. In this case, one can make a prediction about the state of a galaxy based on its $v_1$, which we denote as $f_q(v_1)$. This will on average be a better prediction than $\bar{q}$. As a result, the variance of the corresponding $\Delta$ is reduced relative to the previous case. However, since the hidden variable $v_2$ is spatially correlated, the remaining deviation of $q$ from $f_q$ is still spatially correlated, and the correlation function of $\Delta$ remains unchanged.

Similarly, the bottom-left panel reflects the case in which only $v_2$ is known. For the same reasons as above, the variance is reduced relative to the \lq complete ignorance\rq\ case. However, $f_q$ now accounts for the effects of the correlated $v_2$, and so the spatial correlation of $f_q$ directly follows that of $q$. As a result, the corresponding $\Delta$ is not spatially correlated.

Finally, the bottom-right panel reflects the state of complete knowledge, in which both $v_1$ and $v_2$ are known. In this case, the best prediction that one can make based on $\lbrace v_1,v_2\rbrace$ is identical to the actual mapping of $\mathbf{v}$ to $q$, and therefore $\Delta$ is identically equal to zero. As a result, both $\xi_{\Delta,0}$ and $\sigma_{\Delta}^2$ are zero. 

Overall, the variance behaves as described before, being at its highest in the case of complete ignorance, and monotonically decreasing to zero as our state of knowledge improves. Similarly for the correlation function, when the effect of the correlated variable is accounted for, $\xi_{\Delta}$ vanishes, whereas when only the uncorrelated variable is accounted for, $\xi_{\Delta}$ remains unchanged. 

Since $\gamma_{\Delta}$ expresses the relative importance to quenching of correlated and uncorrelated variables, we would like to better understand its quantitative behaviour. For the case where $\Delta=q-\bar{q}$, $\gamma_{\Delta}$ can be expressed as a function of $\theta$ via \begin{equation} \label{eq:gamma_theta}
\gamma_{\Delta}(\theta) = \int\limits_{-\infty}^{\infty} p(v_1)\,\mathrm{Erf} \left[\frac{\mathrm{tan}(\theta)\,v_1}{\sqrt{2}}\right]^2\ \mathrm{d}v_1,
\end{equation} where $p(v_1)$ is the probability density function of $v_1$, and $\mathrm{Erf}$ is the error function. The graph of this relation is shown in Figure \ref{fig:toy_model_gamma}. From the form of equation (\ref{eq:gamma_theta}), it is clear that even for a simple toy model, $\gamma_{\Delta}$ does not have a simple analytical form. Notice in particular that at $\theta=\pi/4$, where $v_1$ and $v_2$ are given equal importance to quenching, $\gamma_{\Delta}=1/3$ and not $1/2$. For a general $f_q$, finding the analytical form of $\gamma$ is not trivial. Moreover, any deeper analysis of this toy model will probably lose its relevance when compared to real data, where the dimensionality of \vobs\ is higher, where there may be correlations between variables, and where $q$ might not be described by a simple function of $\mathbf{v}$. Therefore, we will simply use the $\gamma_{\Delta}-\theta$ relation here as a qualitative point of reference which provides us with an intuitive sense of the magnitude of $\gamma_{\Delta}$, and therefore an intuitive sense of the relative importance of spatially correlated and uncorrelated variables to quenching.

\subsection{Remarks}
\label{subsec:concept_remarks}

Given the novelty of this conceptual framework, there are a few subtleties which are worth pointing out explicitly.

The first point to note is that the reasoning presented in this Section is fundamentally phenomenological and data-driven. Notice that, in equation (\ref{eq:quenching}), we speak of the variables $\mathbf{v}$ which cause quenching. On the other hand, when modelling $f_q$ in practice, we speak of the variables \vobs\ which are correlated with quenching. The tilde over \lq\vobs\rq\ is intended to indicate that these variables may in fact only be observational proxies of $\mathbf{v}$. For instance, as a result of the known relation of \lq mass quenching\rq, $f_q$ is often modelled as a function of $\mathrm{M}_*$. This is despite the fact that $\mathrm{M}_*$ itself may not be causally responsible for quenching, but rather correlated with a driver of what we refer to as mass quenching. When constructing $f_q$ in practice, we will include variables in the set of \vobs\ on the basis of both correlation with $q$, and our ability to systematically observe them; the fact that they have an observable correlation with $q$ should not be interpreted as a causal link to quenching without physical justification.

A closely related point pertains to our attempt to infer the importance of variables to quenching, based on their ability to account for the variance and the spatial correlation of $\Delta$. In reality, variables which are physically critical to quenching may not necessarily account for very much of the variance or spatial correlation of $\Delta$, and vice versa for variables which are physically unimportant to quenching (see e.g. \citet{Lilly2016} for a study of how this may arise in the case of $\mathrm{M}_*$ and effective radius $R_e$). Therefore, while we will at times discuss the \lq importance\rq\ of variables, these statements should not be thought of as statements about the physical importance of the variables to quenching, but rather as statements about their ability to empirically predict the star-formation states of galaxies.

A third point to note is our assertion that as our state of knowledge becomes more complete, $f_q(\vobs)$ tends toward an exact prediction of $q$. However, $f_q$ is ultimately only a non-parametric estimate, and is based on \vobs, which are practically constrained to be \lq global\rq\ and observable properties of galaxies. On the other hand, it is known that the global star-formation state reflects an average over smaller-scale phenomena across an entire galaxy (e.g. the properties of individual molecular gas clouds). Moreover, it may well be the case that there are drivers of quenching which simply cannot be observed, and have no observable proxy. Our assertion implicitly assumes that the global star-formation state of a galaxy is completely determined by its global, observable properties. If this turns out not to be the case, it is possible that there does not exist an $f_q(\mathbf{\vobs})$ which perfectly predicts $q$. This possibility represents yet another fundamental limitation of our framework.

A last point to elaborate upon is that the \lq hidden variable field\rq\ approach to conformity presented in Section \ref{subsec:conformity_meaning} is a clear departure from how conformity is typically discussed in the literature. At first sight, it may appear strange that key physical properties related to conformity, such as halo membership and the central-satellite distinction, are not built into this framework. This is a result of the fact that we speak generally of variables which are related to quenching. In this sense, halo membership (and the associated $\mathrm{M}_h$) and central-satellite status can be thought of as specific instances of these variables. While they are not built-in aspects of our framework, one can easily include them in the modelling. Specifically, one can mimic a \lq classical\rq\ conformity analysis by constructing $f_{q}(\mathrm{M}_h)$ separately for centrals and for satellites, and measure the correlation of $\Delta$ between these two types of galaxies (i.e. the subsample cross-correlation).

In this work, we opt not to consider halo membership or the central-satellite status in our analysis. The primary reason for doing so is that the determination of these properties rely heavily on the halo memberships presented in group catalogues. As a result, in order to rigorously measure the statistical properties of $\Delta$ constructed around these variables, one necessarily has to take great care in estimating and propagating the systematic errors introduced by errors in group catalogues \citep[see][]{Campbell2015,Treyer2018}. Since the primary motivation of this work is to introduce our framework and methodology, we reserve a more detailed, halo-based study for a future work. Instead, in this work, we account for some of the known dependencies of quenching by modelling $f_q$ as a function of two relatively systematics-free variables, stellar mass $\mathrm{M_{*}}$, and local galaxy number density $\delta$, without distinction between centrals and satellites.

\subsection{Summary of conceptual framework}
\label{subsec:concept_summary}

In this Section, we have developed a conceptual framework to measure and interpret the statistics of the star-formation states of galaxies. Before proceeding, we summarize the key points of this framework.

The key statistic that we have introduced, $\Delta$ (equation (\ref{eq:delta})), is defined as the difference between the star-formation state of a galaxy, $q$, and our best prediction of its state based on our current empirical understanding of galaxy evolution, $f_q$. $\Delta$ can be thought of as a residual which reflects the effects of drivers of quenching which we have not been able to account for via $f_q$ (\lq hidden variables\rq). As $\Delta$ quantifies the incompleteness of our understanding, its variance $\sigma_{\Delta}^2$ summarizes our overall ignorance of the drivers of quenching in a sample. As our state of knowledge becomes more complete, $\sigma_{\Delta}^2$ tends toward zero. We then argued that conformity can be thought of more generally as the spatial correlation of $\Delta$, driven by spatially correlated hidden variables. This can be quantified by the spatial correlation function, $\xi_{\Delta}$ (equation (\ref{eq:corrfun})). Similarly, as our state of knowledge of spatially correlated hidden variables becomes more complete, $\xi_{\Delta}$ tends toward zero.

It then follows that the ratio between $\xi_{\Delta}$ and $\sigma_{\Delta}^2$, which we denote as $\gamma_{\Delta}$, reflects the importance to quenching of spatially correlated (i.e. environmental) variables relative to uncorrelated (i.e. internal) ones. We explored the quantitative behaviour of these statistics within a simple toy model. Although qualitatively, theses statistics behave as expected, the value of $\gamma_{\Delta}$ cannot be expressed in an intuitive way. Therefore, we will simply use the results from the toy model (in particular, Figure \ref{fig:toy_model_gamma}) as a rough guide to interpret the results which we will obtain from the data.

There are a few subtleties to this way of thinking. The first noteworthy point is that the reasoning underpinning this framework is fundamentally phenomenological. The fact that $f_q$ varies with some observables should not lead to the interpretation that these observables are causally related to quenching without physical justification. A related point is that, while we do discuss the empirical importance of variables in terms of their ability to account for the variance and spatial correlation of $\Delta$, these statements should not be interpreted as statements about their physical importance to quenching. A third point to note is that, throughout our reasoning, there is an implicit assumption that the global star-formation state of a galaxy is completely determined by its global properties; if this turns out not to be the case, this framework would require further development. Finally, one should note that the concepts of halo membership and central-satellite status are not built into our framework. Rather, we think of them as specific instances of the more general idea of hidden variables.

In the following Section \ref{sec:data}, we describe the datasets that we will use for analysis. Then, in Section \ref{sec:meth}, we will describe in detail how one can measure the relevant statistics of $\Delta$.

\section{Observational and simulated data}
\label{sec:data}

\subsection{Observational data}
\label{subsec:obs_data}

We use the galaxy sample presented in the New York University Value-Added Galaxy Catalogue\footnote{http://cosmo.nyu.edu/blanton/vagc/} \citep{Blanton2005}, which was constructed with data from Data Release 7 of the Sloan Digital Sky Survey \citep[SDSS DR7;][]{Abazajian2009}. Estimates of stellar masses and star-formation rates are an updated version of those derived in \citet{Brinchmann2004}\footnote{http://wwwmpa.mpa-garching.mpg.de/SDSS/DR7/}. In this work, a galaxy is labelled as quenched if it satisfies $\log(\mathrm{SSFR} [\mathrm{yr}^{-1}])\textless-0.45\log(\mathrm{M}_* [\mathrm{M}_{\odot}])-6.35$.

From this sample, we select galaxies within the redshift range $0.02\,\textless\,z\,\textless\,0.08$, and with stellar masses $\mathrm{M_*}\,\textgreater\,10^9\Msun$. We then compute the mass-completeness limit as a function of redshift. We do so by calculating, for every galaxy in the sample, the redshift at which it would become dimmer than the SDSS magnitude limit of $r=17.7$. We then define the completeness limit at a given mass, $z_{\mathrm{lim.}}(\mathrm{M_*})$, to be the redshift at which the dimmest $5$-percentile of galaxies of that mass would be lost. All galaxies which lie below this $\mathrm{M_*}-z_{\mathrm{lim.}}$ relation are discarded, leaving only a mass-complete sample. 

As a final step in selection, we manually mask regions in the sky where there are holes in the sky coverage of SDSS. The final sample consists of $1\times 10^5$ galaxies.

Our analysis will involve the two-point statistics of galaxies. For this purpose, a useful data structure to consider is a list of pairs of galaxies. Two galaxies are defined to be in a pair if their apparent radial velocities, $v = cz$, differ by no more than $500\,\mathrm{km\,s^{-1}}$. The apparent comoving separation of a pair is defined as $R_{\mathrm{proj.}} = \chi(\bar{z})\Delta\theta $, where the comoving distance $\chi$ is evaluated at the average redshift of the two galaxies, $\bar{z}$. We construct this list to be symmetric, in the sense that if two galaxies, labelled $A$ and $B$, are in a pair, then the list contains both the elements $\{A,B\}$ and $\{B,A\}$. In Section \ref{subsec:meth_caution}, it will be made clear why this property is useful. When defined in this way, our sample contains $6\times 10^6$ pairs with $R_{\mathrm{proj.}}\,\textless\,10\,\mathrm{Mpc}$.

We use the $5^{th}$-nearest-neighbour method to estimate the local galaxy number density, $\delta$. Due to the fact that the mass-completeness limit increases with redshift, the apparent number density of galaxies in our sample decreases with redshift. We define the tracer mass limit as the minimum stellar mass which is complete across our entire redshift range; this mass limit is $\mathrm{M_*}=10^{10.4}\Msun$. In order to avoid this spurious redshift dependence, only galaxies which have masses above this limit are used as tracers of the density field. Also, each sector covered by the SDSS has an associated spectroscopic completeness rate, the inverse of which is used as a weight for the tracers in the corresponding sector.

For each galaxy in our sample, we search for their neighbouring tracers within the list of pairs that we have constructed, and record the sum of the weights of its five nearest tracers, denoted as $W_5$, and the projected comoving separation of its $5^{th}$-nearest-tracer, denoted as $R_{p,5}$. The density estimate, $\delta$, is defined as the weighted number density of tracers in the cylindrical volume within which the tracers are found. That is, \begin{equation} \label{eq:delta} \delta = \frac{W_5}{\pi\, {R_{p,5}}^2\,l}\ ,
\end{equation} where $l=\chi(z+ \frac{\Delta v}{c})-\chi(z- \frac{\Delta v}{c})$, and $\Delta v$ is the $500\,\mathrm{km\,s^{-1}}$ limit used in the construction of the list of pairs.

For galaxies which are near the edges of sky coverage (including masks), or those near the redshift selection limits, the part of this cylindrical volume which lies beyond the selection volume may in reality contain more tracers. For these galaxies, this method will in general produce an estimate which is biased low. As a first order correction for galaxies near edges or masks, we correct the area term, $\pi{R_{p,5}}^2$, by subtracting from it the area of the part of the circle which lies outside of coverage. Similarly, for those near the redshift limits, we subtract from the length term, $l$, the portion of it which lies outside of the limits.

Finally, we will also make reference to the group membership of SDSS galaxies at some points in the discussion. We do so by making use of the Yang et al. SDSS DR7 group catalogue\footnote{http://gax.shao.ac.cn/data/Group.html}, the construction of which is described in \citet{Yang2007}.

\subsection{Mock data}
\label{subsec:mock_data}

In order to compare the results from observations against those predicted by galaxy formation models, we also use the semi-analytic model (SAM) of \citet[][hereafter H15]{Henriques2015}\footnote{http://galformod.mpa-garching.mpg.de/public/LGalaxies/}. This is the most recent major release of the so-called Munich models and was implemented on the Millennium dark matter simulations scaled to a Planck-year1 cosmology \citep{Planck2014}. Specifically, the cosmological parameters adopted are: $\sigma_8 = 0.829$, $H_0 = 67.3 \,\mathrm{km\,s^{-1}Mpc^{-1}}$, $\Omega_{\Lambda} = 0.685$, $\Omega_{\mathrm{M}} = 0.315$, $\Omega_{\mathrm{b}} = 0.0487$ $(f_{\mathrm{b}} = 0.155)$ and 
$n = 0.96$. We use the galaxy catalogue based on the Millennium simulation since it has a larger volume (meaning better statistics for satellite galaxies), and H15 showed that its properties converge with those in the catalogue based on the higher-resolution Millennium-II down to our low-mass limit ($M_*=10^{9} \Msun$). Both the Millennium and Millennium-II simulations trace $2160^3$ ($\sim10$ billion) particles from $z = 127$ to the present day. The Millennium was carried out in a box of original side $500\,h^{-1} \mathrm{Mpc} = 685 \, \mathrm{Mpc}$. After rescaling to the Planck cosmology, the box size becomes $714\,\mathrm{Mpc}$, implying a particle mass of $1.43\times10^9 \Msun$.

From the $z=0$ snapshot of the output, we construct a mock catalogue which is similar to the data. We first convert the full 6-dimensional position-velocity data into 3-dimensional observed coordinates ($x$, $y$, redshift) by converting the position and velocity along one direction into redshift, omitting the velocities along the other two directions. We then apply the same redshift, mass, and completeness cuts as those used for the observational data, and construct the list of pairs and the density estimates in the same way. This results in $8\times 10^5$ galaxies in the mock, and $7\times 10^7$ pairs with $R_{\mathrm{proj.}}\,\textless\,10\,\mathrm{Mpc}$.

One difference to note is that whereas the geometry of the SDSS is conical, the geometry of our mock data is cubic. This, in addition to the redshift-dependent completeness cut, will cause the distribution of galaxies in $(\mathrm{M_{*}},\delta)$-space to somewhat differ between the SDSS and the mock. Another difference to note is that, given the completeness of information in the mock, the identification of groups in the mock is far less prone to error than that in the observational data.

\section{Methodology}
\label{sec:meth}

\subsection{The estimation of $\boldsymbol{f_q(\mathrm{M_{*}},\delta)}$}
\label{subsec:fq}

In Section \ref{subsec:fq_meaning}, we discussed the conceptual significance of fitting $f_q(\vobs)$. In practice, controlling for a large number of variables is infeasible for several reasons. First, as discussed before, it is likely the case that some variables which are relevant for quenching have not been systematically observed for our sample. Therefore, even when there is theoretical motivation to model the effects of a variable \citep[e.g. assembly bias,][]{Hearin2016}, reliable observable proxies may not be readily available. Second, as the number of variables considered in the modelling of $f_q$ increases, the number density of data points in variable space is diluted exponentially, which limits the accuracy of the model.

In this work, we will control for two variables which are systematically observed, and are known to have independent correlations with quenching: the stellar mass $\mathrm{M_*}$, and the local environment of galaxies $\delta$ \citep{Peng2010}. First, for illustration, we consider the effects of these two variables separately, which we do by fitting $f_q$ as a function of either $\mathrm{M_*}$ or $\delta$. Given a sample of galaxies (this could be either the SDSS data, or the mock data), we estimate the $f_q$ of a given galaxy in the sample by taking the average $q$ of its neighbours within a given \lq smoothing length\rq\ in variable space. For the case of estimating $f_q(\mathrm{M_*})$, we use a smoothing length of 0.1 dex in $\mathrm{M_*}$-space (i.e. $\pm 0.05\,\mathrm{dex}$), while for estimating $f_q(\delta)$, we use a smoothing length of 0.15 dex in $\delta$-space. We note however that the particular choices of smoothing length, or even fitting method, has relatively little impact on the corresponding results.

Then, we account for the effects of both variables simultaneously by setting $f_q$ to be $f_q(\mathrm{M_{*}},\delta)$. We estimate $f_q(\mathrm{M_{*}},\delta)$ for every galaxy in a sample (again, this could be either the SDSS data or the mock) by taking the average $q$ of its nearest-neighbours in $(\mathrm{M_{*}},\delta)$-space. That is to say, given a galaxy $i$ and the set of its $k$-nearest neighbours $\mathbf{N}_i$, the estimator for $f_q(\mathrm{M_{*}},\delta)$ is defined as \begin{equation} \label{eq:fq_est} f_{q,i}= \frac{1}{k}\sum_{j \in\mathbf{N}_i} q_j.
\end{equation} In order to do so, one must define a distance measure in variable space, and determine a suitable number of nearest-neighbours over which to take the average.

We define the distance between two points in variable space separated\footnote{For the cleanliness of notation, it is implicit that separations in variable space, $\Delta\mathrm{M_*}$ and $\Delta\delta$, are in fact shorthands for the differences in the logarithms, $\Delta\mathrm{log}(\mathrm{M_*}\,[\mathrm{M}_{\odot}])$ and $\Delta\mathrm{log}(\delta\,[\mathrm{Mpc^{-3}}])$.} by ($\Delta\mathrm{M_*}$,$\Delta\delta$) as \begin{equation} \label{eq:pspace_dist}
D_{\mathrm{M_{*}},\delta}=\sqrt{(1.5\,\Delta\mathrm{M_*})^2 + (\Delta\delta)^2},
\end{equation} where the factor of 1.5 scales $\mathrm{M_*}$ such that the width of the distribution of $\mathrm{M_*}$ (in dex) is roughly equal to that of the distribution of $\delta$. 

We optimize the number of variable-space neighbours over which to take the average, denoted by $k$, on a per-galaxy basis. On one hand, estimating $f_{q}$ using a small number of points would be sensitive to statistical fluctuations in $q$. On the other hand, using a large number of points necessarily biases the estimate towards the sample average. We would like to choose $k$ such that the effects of variance and bias are balanced. However, the $k$ at which these two effects are balanced depends on the local distribution of points in variable space, and therefore we optimize $k$ on an per-galaxy basis. For the same reason, we optimize $k$ separately for the observational data and for the mock. 

We optimize $k$ by evaluating the accuracy of the estimator of $f_q$ (equation (\ref{eq:fq_est})) on a fiducial $f_{q,\mathrm{fid.}}$ for a range of $k$. We first assign to each galaxy in the sample a fiducial $f_{q,\mathrm{fid.}}(\mathrm{M_{*}},\delta)$ which roughly resembles that which is observed in the data. From this $f_{q,\mathrm{fid.}}$, we stochastically generate 100 independent realizations of the star-formation states (i.e. $f_{q,\mathrm{fid.}}$ defines the probability of $q$ being 1, and $1-f_{q,\mathrm{fid.}}$ is the probability of $q$ being 0). This random generation of $q$ is equivalent to having spatially uncorrelated variables which are hidden relative to $(\mathrm{M_{*}},\delta)$. Then, for each galaxy, $\hat{f}_q$ is estimated for each realization for a range\footnote{$k$ is allowed to vary from 100 to 1500 for the data, and from 100 to 3000 for the mock. The lower limit of 100 is set such that $f_q$ is not overly quantized, while the upper limits are heuristic thresholds at which increasing $k$ offers diminishing returns in reducing the variance.} of $k$. The many realizations allow us to estimate the bias $b = \langle f_{q,k} - f_{q,\mathrm{fid.}}\rangle$, as well as the effects of shot noise via the variance $\sigma^2 = \mathrm{Var}\left[f_{q,k}\right]$. The optimal $k$ for a galaxy is then defined as the value at which $b^2 + \sigma^2$ is minimal.

Given an optimal $k$ for each galaxy, $f_q(\mathrm{M_{*}},\delta)$ is then calculated via equation (\ref{eq:fq_est}). Using this method on the SAM results in estimates of $f_q$ for which $\langle b\rangle \approx 0$, $\langle \sigma\rangle \approx 0.01$, and both with spread of order $10^{-3}$. Similar results are achieved for the data, with a roughly factor of 2 larger spread due to the smaller size of the data set. Needless to say, this also implies that $\Delta$ can be estimated to the same order of accuracy and precision. The propagation of the systematic errors of $f_q$ to the quantities of interest, $\sigma_{\Delta}^2$ and $\xi_{\Delta}$, can also be estimated by making use of the many realizations of $q$. We do so by first evaluating, for each realization, the $\Delta$ of every galaxy in a sample by substituting $f_q$ into equation (\ref{eq:delta}), and then evaluating $\sigma_{\Delta}^2$ and $\xi_{\Delta}$. These quantities can then be compared to their respective fiducial values. The fiducial variance is $\langle f_{q,\mathrm{fid.}}(1-f_{q,\mathrm{fid.}})\rangle$, and since $q$ is generated randomly based solely on $f_{q,\mathrm{fid.}}$, there is no conformity in the simulated data, and so the fiducial $\xi_{\Delta}$ is zero at all $R$. In doing so for both the SDSS data and the mock, we find that the average deviation of $\sigma_{\Delta}^2$ and $\xi_{\Delta}(R)$ from their respective fiducial values are consistent with zero, with spread of order $10^{-4}$.

\subsection{The measurement of $\xi_{\Delta}(R)$}
\label{subsec:corr_func}

In this subsection, we discuss how $\Delta$ can be used as an indicator of conformity. Within our framework, the spatial correlation of $\Delta$ is the indicator of conformity. This correlation can be measured via the spatial correlation function $\xi_{\Delta}(R)$, given in equation (\ref{eq:corrfun}).

In practice, the correlation of the $\Delta$ field can only be measured from a finite set of galaxies, for which only their projected positions on the sky, $\hat{\mathbf{x}}$, can be observed. Correspondingly, one can define the projected correlation function \begin{equation} \label{eq:obs_corrfun}
w_{f}(R_p)=\langle f(\hat{\mathbf{x}})f(\hat{\mathbf{x}}\rq)\rangle,
\end{equation} where the angled brackets denote an average which is taken over all pairs with $\left|\hat{\mathbf{x}}-\hat{\mathbf{x}}\rq \right|=R_p$.

Under a projection from a 3-D real-space distribution to 2-D observed positions on the sky, pairs of galaxies at a given projected separation could have a range of real-space separations. In fact, at a given projected separation $R_{p}$, the magnitude of the real-space separation $\left| R\,\right|$ is at least $\left| R_{p}\right|$, but possibly much larger. The effect that this has on the projected correlation can be expressed as the convolution \begin{equation} \label{eq:conv_integral}
w_{f}(R_{\mathrm{p}})=\int_{0}^{\infty}p(R\,\vert\,R_{p})\,\xi_{f}(R)\ dR,
\end{equation} where $p(R\,\vert\, R_{p})$ denotes the normalized distribution of real-space separations $R$ at a given $R_{p}$, and will naturally depend on the way in which pairs are selected. Supposing that $\xi_{f}$ decreases with increasing $R$, the effects of equation (\ref{eq:conv_integral}) would be to suppress $w_{f}$ relative to $\xi_{f}$. 

We would like to recover an estimate of $\xi_{f}$ from $w_{f}$. This is possible for the mock, where information about both $R$ and $R_p$ is available. We do so by limiting ourselves to a finite range of separations, discretized as $\mathbf{R}_{p}=\left[R_{p,1},\hdots,R_{p,M}\right]$, $\mathbf{R}=\left[R_{1},\hdots,R_{N}\right]$. Under this discretization, equation (\ref{eq:conv_integral}) can be approximated as the matrix equation 
\begin{multline} \label{eq:conv_matmul1}
\left[ \begin{array}{c} w_{f}(R_{p,1}) \\ \vdots \\ w_{f}(R_{p,M})   \end{array} \right]
= \\
\begin{bmatrix} 
p(R_{1}\vert R_{p,1}) & \hdots & p(R_{N}\vert R_{p,1}) \\
\vdots & \ddots & \vdots \\
p(R_{1}\vert R_{p,M}) & \hdots & p(R_{N}\vert R_{p,M}) \end{bmatrix}
\left[ \begin{array}{c} \xi_{f}(R_{1}) \\ \vdots \\ \xi_{f}(R_{N})   \end{array} \right],
\end{multline} or more compactly, \begin{equation} \label{eq:conv_matmul2}
\boldsymbol{w}_{f}=\boldsymbol{\mathcal{C}}\,\boldsymbol{\xi}_{f},
\end{equation} where the $i^{th}$ row of the matrix $\boldsymbol{\mathcal{C}}$ is the normalized and discretized distribution of $R$ at $R_{p,i}$. Such an approximation is valid if $p(R\,\vert\,R_{p})$ tends to zero at $R_N$, which is a physically justified assumption if $R_N$ is chosen to be sufficiently large. 

In order to deconvolve the observed signal, one must therefore consider the convolution matrix $\boldsymbol{\mathcal{C}}$. While $\boldsymbol{\mathcal{C}}$ is not known for the data, it can easily be constructed for the mock. For simplicity, the dimension of $\mathbf{R}$ is chosen to be the same as that of $\mathbf{R}_p$ (i.e. $M=N$), such that $\boldsymbol{\mathcal{C}}$ is square, and therefore invertible\footnote{In order to be invertible, the columns of $\boldsymbol{\mathcal{C}}$ are also required to be linearly independent. By the physical nature of $p(R\,\vert\, R_{p})$, $\boldsymbol{\mathcal{C}}$ is upper-triangular with non-zero elements on the diagonal, and so this condition is also met.}. Given the above, the deconvolution can be expressed as \begin{equation} \label{eq:deconv}
\boldsymbol{\xi}_{f}=\boldsymbol{\mathcal{C}}^{-1}\boldsymbol{w}_{f}.
\end{equation}

\begin{figure}
\centering
\includegraphics[width=8.6cm]{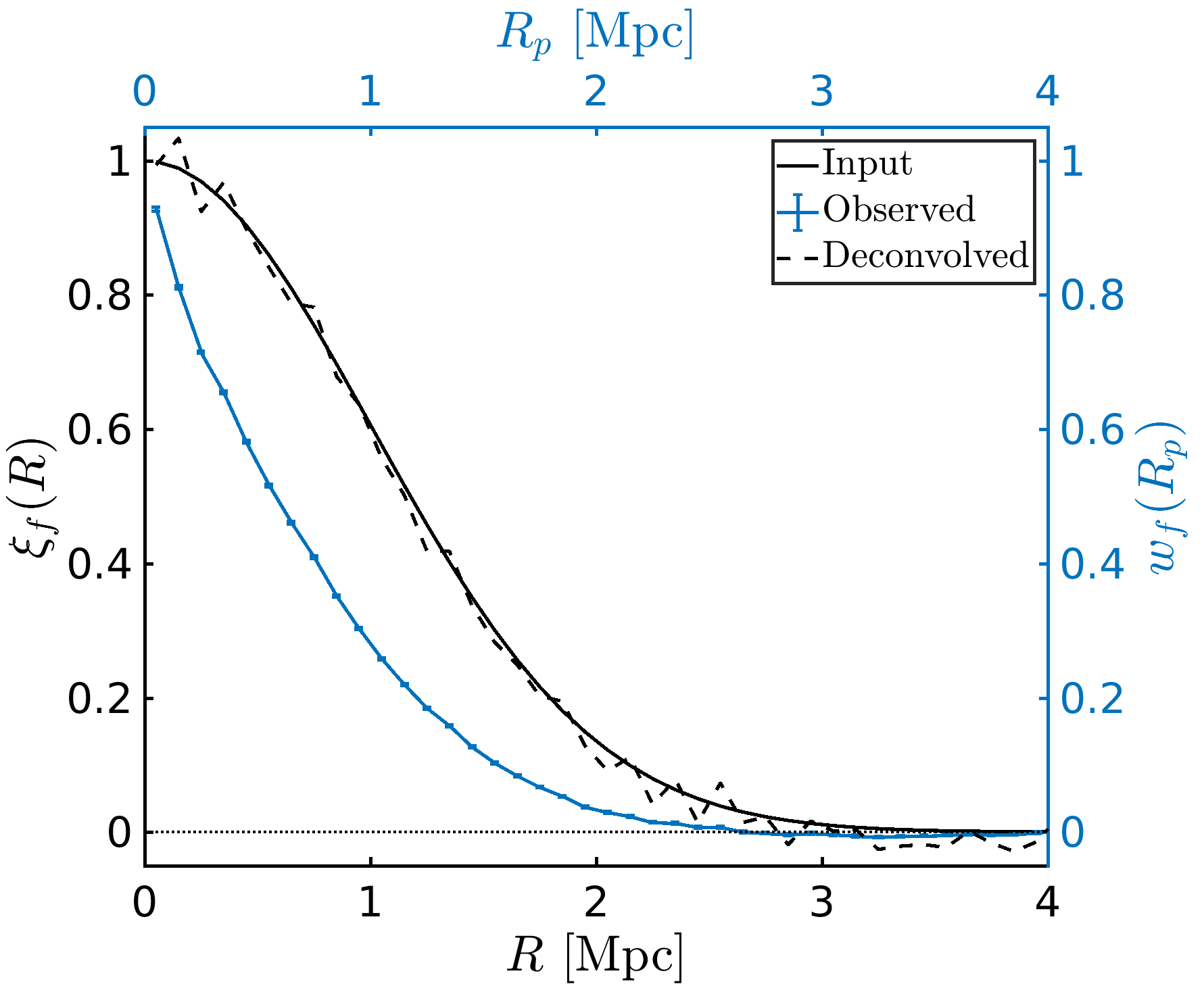}
\caption{A test of our ability to constrain the correlation of a field in the mock under projection. Solid black: The input correlation function, defined in terms of the real-space positions of the galaxies. Blue: The measured correlation function, measured in terms of the projected separations of pairs of galaxies. The errorbars indicate the standard error on the mean. Dashed black: The result of deconvolving the measured signal, which corrects for the effects of projection. The result mostly overlaps with the original input signal.}
\label{fig:sam_cftest}
\end{figure}

The ability to easily construct $\boldsymbol{\mathcal{C}}$ for the mock allows us to test the efficacy of this deconvolution scheme. We do so by first generating a Gaussian random field $f$ in the simulation volume, with $\mu_f=0$, $\sigma_f^2=1$, and a spatial correlation based on the full 3-D positions of the mock galaxies, $\xi_f(R) = \sigma_{f}^2\,\mathrm{exp}(-R^2/2)$. We then measure $w_{f}$ from the projected positions of the galaxies, and deconvolve the signal as described above. The results are shown in Figure \ref{fig:sam_cftest}. The solid black line shows the form of the input correlation function, while the blue line and points show the measured signal. As expected, $w_{f}$ is suppressed relative to $\xi_f$. The dashed black line shows the result of the deconvolution; while somewhat noisy, the deconvolution recovers the input signal accurately. 

In this work, we will also assume that the $p(R\,\vert\, R_{p})$ of the mock is a sufficiently good approximation of that of the SDSS sample, and therefore use the matrix $\boldsymbol{\mathcal{C}}$ constructed from the mock to deconvolve the signal from the observational data. One should keep in mind that the signal measured from the observational data will be noisier due to the smaller sample size. For these reasons, although we will estimate $\xi_{\Delta}$ for the observational data, one should only treat it as a rough estimate.

\subsection{Methodological considerations}
\label{subsec:meth_caution}
When using pairs of galaxies as probes of a signal, there are several methodological issues to consider when interpreting the resulting signal. Some of these were discussed in detail in an earlier work of ours \citep{Sin2017}. While some of these issues were relevant only to the methodology of \cite{Kauffmann2013}, others are more generally applicable. In this subsection, we provide a brief review of two issues which are relevant to our analysis, and possible methods to address them.

\begin{enumerate}
\item \textbf{Bias due to density-weighting}: When averaging over pairs, as required in equation (\ref{eq:obs_corrfun}), every pair within the sample is implicitly given equal weight. As a consequence, a region with number density $n$, which would produce on the order $n^2$ pairs, would have an implicit weight of order $n^2$. That is to say, high-density regions would drastically outweigh low-density regions. If conformity were environment-dependent, the signal from high-density regions would overwhelm that from the remainder of the sample.

In order to address this issue, one can down-weight galaxy pairs in proportion to their local galaxy number density. Operationally, we implement this as follows: if galaxy $A$ appears as the first element of the pair in $N_A$ pairs, those pairs are down-weighted by $1/N_A$. Such a weighting scheme approximately reduces the density-bias from an order of $n^2$ to an order of $n$; roughly speaking, this gives every galaxy equal weight\footnote{It is for this purpose that the list of pairs is constructed to be symmetric (see Section \ref{subsec:obs_data}). Consider two galaxies, $A$ and $B$, where $A$ is paired with many galaxies, but $B$ is only paired with $A$. If element $\lbrace A,B\rbrace$ were in the list of pairs, but $\lbrace B,A\rbrace$ were not, then this weighting scheme would effectively down-weight $B$ by as much as $A$, which defeats the purpose of the weighting. It is only when $\lbrace B,A\rbrace$ exists that $B$ would remain at weight 1, and thereby achieve our desired effect.}. Furthermore, if one wished to give every unit volume equal weight (i.e. with no bias towards $n$), one should in principle use a $1/N^2$ weight. However, since weighting effectively increases the variance of a sample, this in practice results in an excessively noisy measurement for our observational sample. Therefore, only the $1/N$ weighting scheme will be implemented in this work.

\item \textbf{Interpretation of length-scales}: The second issue also arises from the fact that a signal from the full sample will necessarily be probing a range of environments. Specifically, given our sample, the signal will originate from galaxies within halos of a range of masses; the most massive of these haloes have virial diameters on the order of 4 Mpc. As conformity is known to be (at least) a halo-scale phenomenon, one should therefore not be surprised to measure a signal on a correspondingly large scale. More generally, the length-scale of a measured signal should only be interpreted within the context of the environment from which the signal originates.

To address this issue, one needs to first determine the halo membership of galaxies, for instance with reference to a group catalogue. In \cite{Sin2017}, this information was used to mask all pairs for which one or both elements are near large haloes.
\end{enumerate}

By applying these two methodological modifications, \cite{Sin2017} were able to prevent the signal from high-density regions from dominating the overall signal. Moreover, it was found that the \cite{Kauffmann2013} measurement of a large-scale ($4\,\mathrm{Mpc}$) conformity signal originates from galaxies which are within, or associated with, individual massive haloes with virial diameters of up to $4\,\mathrm{Mpc}$, a signal which was then amplified by the density-weighting effect. After accounting for this, there remained no strong evidence for \lq two-halo\rq\ conformity.

Within the current framework, there is an additional possible approach. The fundamental issue underlying (i) and (ii) is essentially that the magnitude and scale of conformity can depend on various properties of the galaxy population, the details of which are lost by averaging over the full list of pairs of the sample. However, it is not necessary to average over all pairs when measuring the correlation of $\Delta$. Rather, one can do so for any subsample of interest. Generally speaking, for any subsample, the associated list of pairs would be a subset of the full list of pairs, in which both elements of a pair are members of the subsample of interest. The corresponding correlation function (i.e. the subsample autocorrelation) is simply equation (\ref{eq:obs_corrfun}) evaluated for this selected list of pairs\footnote{One could also evaluate the cross-correlations of the subsamples. However, we have found that, when evaluated for either the SDSS data or the mock, they do not yield additional useful information, and therefore we exclude them from this work.}.

For example, consider the full sample divided into independent subsamples selected in density. This method, when applied to these subsamples, would reveal the variation of conformity with density, and could be an alternative way to address issue (i).

This method also provides a way to address the one- or two-halo nature of the correlation function. A pair in which both elements are members of the same halo is a probe of the one-halo component of the signal, while a pair containing members of two different haloes would be a probe of the two-halo component. Thus, dividing the full list of pairs accordingly would be a natural way to decompose the full correlation signal into its one- and two-halo components, and could be an alternative way to address issue (ii). This method of decomposing the full-sample signal may yield more insight than the methodological modifications specific to issues (i) and (ii), as it enables us to consider different environments separately, without down-weighting or masking the signal from any particular subset of galaxies.

In this work, we will apply the $1/N$ weighting, as well as evaluate the autocorrelation of subsamples.

\subsection{Summary of methodology}
\label{subsec:meth_summary}

In this Section, we have described the methodology needed to meaningfully measure $\Delta$, and its correlation function. Before proceeding, we summarize the key points of our methodology.

We first described our method for estimating $f_q$ as a function of $\mathrm{M}_*$, $\delta$, and both, from a sample of galaxies. Through rigorous testing, we find that our methods do not introduce statistically significant errors to our desired measurements.

While we would like to measure the real-space correlation function of $\Delta$, one can in practice only measure its 2-D projection, $w_\Delta$. This projection can be approximated as the result of a matrix multiplication (equation (\ref{eq:conv_matmul1})). With the aid of the mock catalogue, one can estimate the convolution matrix $\boldsymbol{\mathcal{C}}$, compute its inverse, and thereby estimate $\xi_\Delta$ from $w_\Delta$. Testing this deconvolution scheme on a fiducial signal in the mock shows that it can accurately recover $\xi_\Delta$ from $w_\Delta$.

There are also methodological aspects which, if not understood, may lead to misinterpretations of the results. The first issue is the \lq density-weighting\rq\ bias, where the implicitly equal weighting of pairs in evaluating equation (\ref{eq:obs_corrfun}) results in a bias towards high-density regions (which naturally contain many pairs). The second issue is the difficulty in interpreting length-scales. As conformity is already known to be a halo-scale phenomenon, one would expect $\Delta$ to be correlated at least on halo scales. However, given that the sample contains haloes with a range of virial radii, the scale of the correlation measured from the full sample should not be interpreted directly, but rather with regard to the diverse environments from which its components originate.

The fundamental issue underlying the above two is that the correlation of $\Delta$ can depend on various properties of the galaxy population. To address this, one can split the full sample into subsamples (e.g. by density), or split the list of pairs into subsets (e.g. into \lq same-halo\rq\ or \lq different-halo\rq\ pairs). The corresponding subsample autocorrelations could then yield further insight into the physical nature of conformity.

\section{Results}
\label{sec:results}

Having developed our methodology, and demonstrated its ability to accurately and meaningfully measure conformity, we now move on to applying it to the SDSS data. We then apply an identical analysis to the mock in order to see what insights we can gain about the physical nature of conformity.

The statistical uncertainties of $\sigma_{\Delta}^2$ and $w_{\Delta}(R_p)$ are both estimated via 100 iterations of bootstrap resampling. In the case of $w_{\Delta}(R_p)$, the resampling of the distribution of $\Delta(\hat{\mathbf{x}})\Delta(\hat{\mathbf{x}}\rq)$ is performed within bins of $R_p$, and the standard deviation of the means of the resampled distributions is taken as the statistical uncertainty of $w_{\Delta}$ at a given $R_p$. In the down-weighted case, the weight of a given pair of galaxies is fixed, such that when the $\Delta(\hat{\mathbf{x}})\Delta(\hat{\mathbf{x}}\rq)$ of a pair is drawn in the resampling, the weight of the pair is drawn along with the value. We will also directly use the uncertainties of $w_{\Delta}$ as the uncertainties of the corresponding deconvolved signal.

\subsection{Results from observational data}
\label{subsec:obs_res}

\begin{figure*}
\centering
\includegraphics[width=17.9cm]{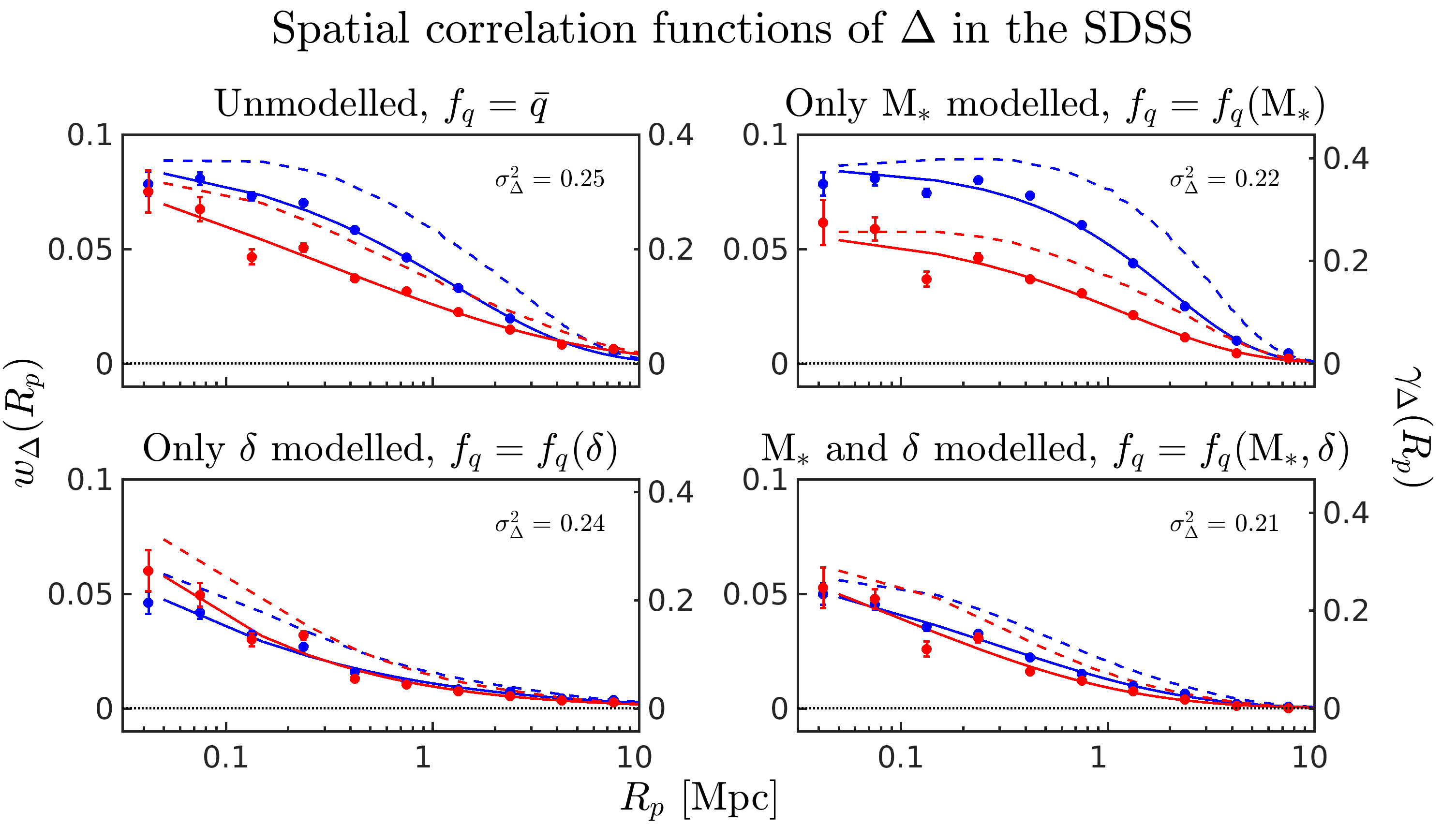}
\caption{Correlation functions of $\Delta$ measured from the SDSS data, where the different panels make use of different $f_q$\rq s to illustrate the impact of different variables on $\Delta$. In the top-left panel, $f_q$ does not model the effects of any variables. In the top-right and bottom-left panels, $f_q$ respectively models the average effects of $\mathrm{M_*}$ and $\delta$. Finally, in the bottom-right panel, $f_q$ models the average effects of both $\mathrm{M_*}$ and $\delta$. The blue points are the results of evaluating equation (\ref{eq:obs_corrfun}) under a weighting scheme where all pairs of galaxies are weighted equally. The red points result from a scheme where pairs of galaxies are down-weighted, such that every galaxy has roughly equal weight. The blue and red solid lines are the results of fitting a function (equation (\ref{eq: signal_fit})) to the respective set of points. The dashed lines show the deconvolved signals, which result from applying the mock\rq s deconvolution matrix to the fitted lines. The text in each panel indicate the variance of the corresponding $\Delta$, while the vertical axes on the right of each panel indicate the corresponding $\gamma_\Delta$.}
\label{fig:sdss_fullcorr}
\end{figure*}

\begin{figure}
\centering
\includegraphics[width=8.6cm]{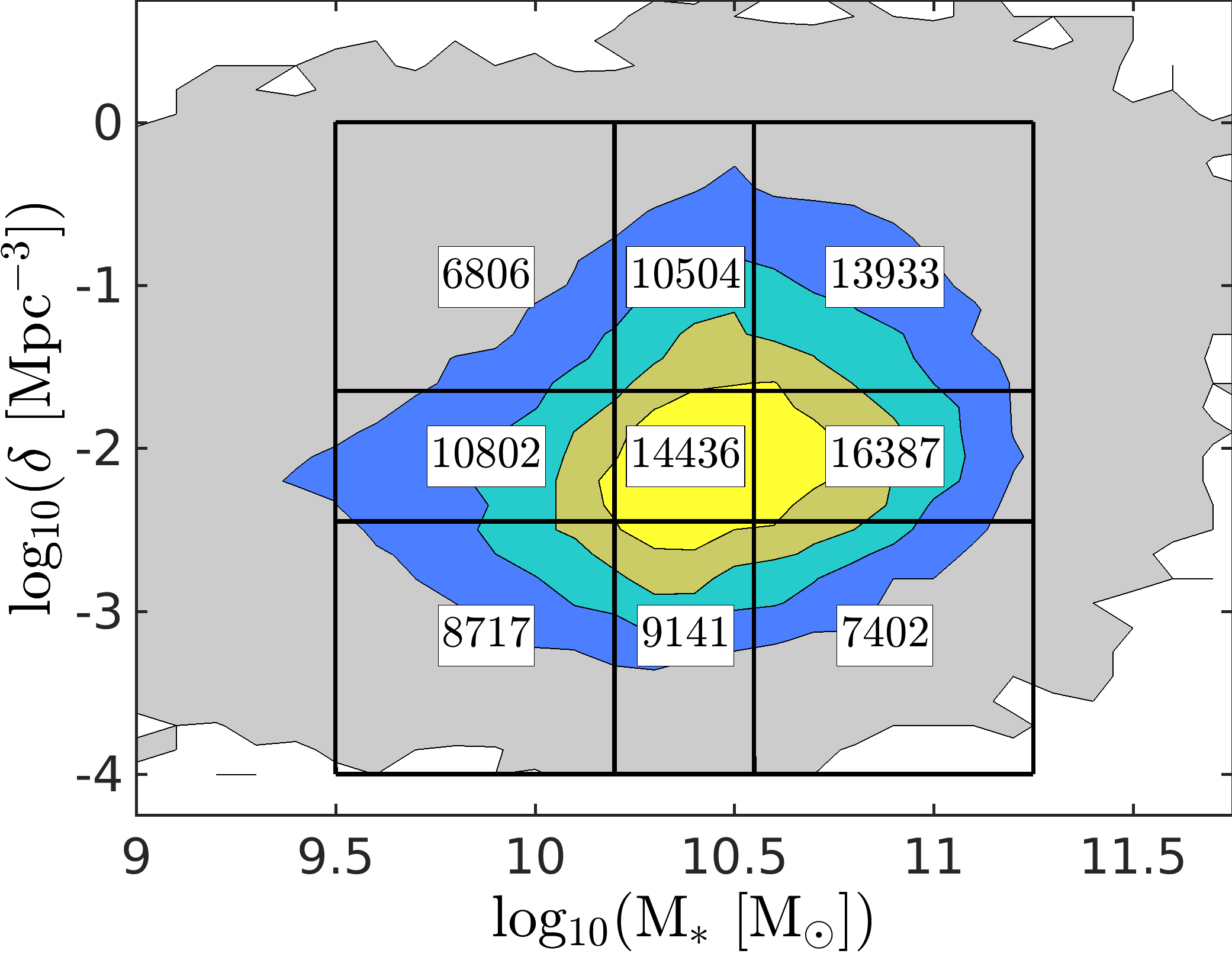}
\caption{Subsamples of the SDSS data selected in $\mathrm{M_*}$ and $\delta$, for the purpose of measuring the dependence of conformity on these variables. The black rectangles indicate the limits of the subsample selections, and the labels indicate the respective subsample sizes. The contours indicate the percentiles of the sample distribution at 20-percent intervals.}
\label{fig:sdss_subsets}
\end{figure}

\begin{figure*}
\centering
\includegraphics[width=17.9cm]{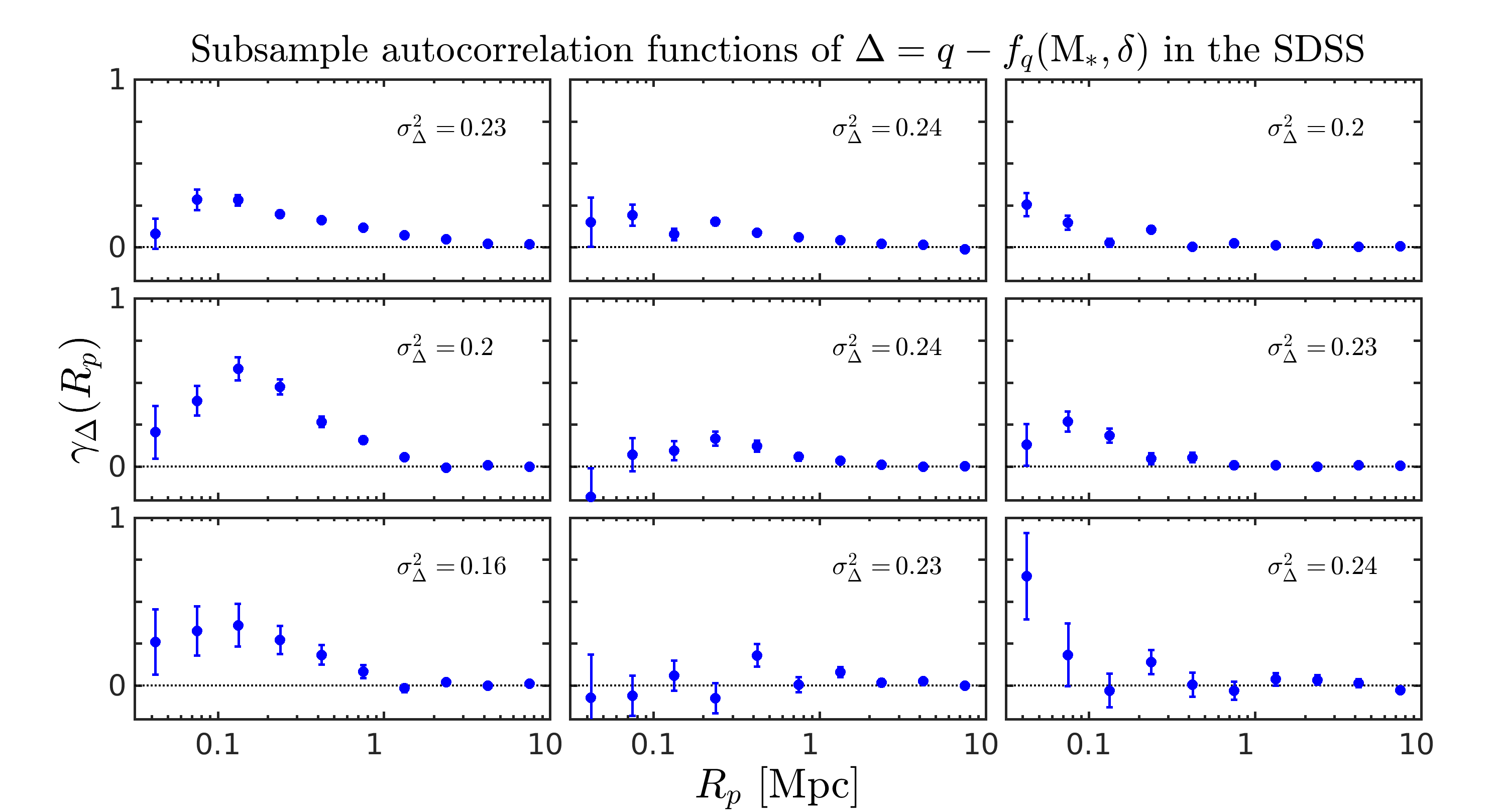}
\caption{The unweighted subsample autocorrelation functions of $\Delta=q-f_q(\mathrm{M_*},\delta)$, measured within the subsamples indicated in Figure \ref{fig:sdss_subsets}. The positions of the panels correspond to the positions of the divisions in Figure \ref{fig:sdss_subsets}, with $\mathrm{M_*}$ increasing from left to right, and $\delta$ increasing from the bottom to the top. The results are plotted as a fraction of the respective subsample variances, which is also displayed in the text within each panel.}
\label{fig:sdss_subcorrs}
\end{figure*}

We first measure, for the SDSS data, the relevant statistical properties of $\Delta$ using different $f_q$\rq s. The four cases we will consider are \begin{enumerate}
\item $f_q=\bar{q}$, which assumes no knowledge about galaxies,
\item $f_q=f_q(\mathrm{M_*})$, which accounts for the average effects of \lq mass quenching\rq,
\item $f_q=f_q(\delta)$, which accounts for the average effects of \lq environmental quenching\rq,
\item $f_q=f_q(\mathrm{M_*},\delta)$, which accounts for the effects of both mass and environment.
\end{enumerate}

In Figure \ref{fig:sdss_fullcorr}, the four panels show the measured correlation function of $\Delta$ for these four cases. Within each panel, the unweighted, projected correlation function $w_{\Delta}$ is plotted in blue points, and the corresponding down-weighted correlation function is plotted in red.

We also attempt to recover $\xi_{\Delta}$ via deconvolution. To avoid the amplification of noise in the signal, we first fit a function of the form \begin{equation} \label{eq: signal_fit}
f(R_p)= a\,\mathrm{exp}(-b\,{R_p}^c) 
\end{equation} to the measured points, where $a$, $b$, and $c$ are free parameters of the fit. We then construct the convolution matrix from the mock, and use it deconvolve the fitted function. As noted in Section \ref{subsec:corr_func}, the geometry of the mock and the SDSS are different, and so the $\boldsymbol{\mathcal{C}}$ constructed from the mock is not expected to be exactly applicable to the data. Therefore, the $\xi_{\Delta}$ presented here should only be thought of as a rough approximation. Also, due to the discrete and \lq empirical\rq\ nature of this deconvolution scheme, we can not estimate $\xi_{\Delta}(R)$ for arbitrarily small values of $R$. Instead, we directly take the value of $\xi_{\Delta}$ at $R=0.005\,\mathrm{Mpc}$ (i.e. the innermost bin of the deconvolved signal) as $\xi_{\Delta,0}$, which we then use for the calculation of $\gamma_\Delta$. The result of deconvolving the unweighted and down-weighted measurements (hereafter, \lq u.w.\rq\ and \lq d.w.\rq) are plotted as blue and red dashed lines, respectively.

As we are interested in the relative importance of spatially correlated and spatially uncorrelated hidden variables, we will also express $w_\Delta$ and $\xi_{\Delta}$ as a fraction of the respective variance in each case; by a slight extension of the original definition of $\gamma_\Delta$ (equation (\ref{eq:coherence_factor})), we denote this fraction as $\gamma_\Delta(R_p)$, and display its value on the right-hand axis of each panel.

For case (i), where $f_q=\bar{q}$, we find that the overall effects of hidden variables lead to $\sigma_{\Delta}^2=0.24974\pm 0.00004$. The spatially correlated hidden variables lead to $\xi_{\Delta,0}=0.089\pm0.005$ (u.w.) and $0.079\pm0.009$ (d.w.), which correspond to $\gamma_\Delta=0.36\pm0.02$ (u.w.) and $0.32\pm0.04$ (d.w.). Within the toy model (see Figure \ref{fig:toy_model_gamma}), these values of $\gamma_\Delta$ roughly correspond to the regime where the internal and external hidden variables have roughly equal importance to quenching. The strength of the correlation decays with increasing separation, but persists out to 10 Mpc.

For case (ii), where $f_q=f_q(\mathrm{M_*})$, we find that the remaining hidden variables lead to $\sigma_{\Delta}^2=0.2240\pm 0.0004$, which is a modest decrease relative to (i). On the other hand, the change in the correlation functions are minor; this is especially true for the down-weighted result, where the change in $w_\Delta(R_p)$ relative to (i) is statistically insignificant. By accounting for the average effects of mass quenching, the total variance of the residual is reduced, while the spatial correlation of $\Delta$ is unaffected. This is unsurprising, since $\mathrm{M}_*$ is a property of individual galaxies, and is mostly not spatially correlated.

For case (iii), where $f_q=f_q(\delta)$, we find that the remaining hidden variables lead to $\sigma_{\Delta}^2=0.2362\pm 0.0003$, which is again a modest decrease relative to (i). On the other hand, for the correlation of $\Delta$, we find $\xi_{\Delta,0}=0.059\pm0.005$ (u.w.) and $0.074\pm0.009$ (d.w.), corresponding to $\gamma_\Delta=0.25\pm0.02$ (u.w.) and $0.31\pm0.04$ (d.w.), which is a notable decrease relative to (i). More remarkable, however, is the reduction in the scale of the spatial correlation of the residual. On the 1-Mpc scale, $w_\Delta$ is approximately half of that in (i). By accounting for the average quenching effects of $\delta$, both the total variance and the spatial correlation of $\Delta$ are reduced. Again, this is as expected, given that $\delta$ is an environmental property.

It is interesting to note that accounting for the effects of environment is not as effective in reducing the variance in $\Delta$ as doing the same for mass. That is to say that, averaging over our sample, the effects of mass quenching play a more important role in quenching than the effects of environmental quenching. It is also worth noting that $\delta$ alone is not a complete description of environmental effects, as there remains a (relatively weak) correlation signal which persists out to 10 Mpc. 

Apart from the change in strength, there is another noteworthy aspect of the behaviour of the correlation of $\Delta$. Note that by accounting for the average quenching effects of the local density $\delta$, the unweighted and the down-weighted correlation functions become essentially the same. To better understand this behaviour, recall from the discussion of issue (i) in Section \ref{subsec:meth_caution}, that the motivation for our density-dependent weighting arises from the fact that regions of high-density produce a large number of pairs relative to those of low-density, and therefore that an unweighted measurement of an environment-dependent effect would be dominated by high-density regions. This effect is obvious in the top-left panel, where the strength of $w_{\Delta}$ depends on the weighting scheme. Specifically, an environment-dependent correlation function of $\Delta$ can arise either because the average $q$ varies with $\delta$, or because the strength of the spatial correlation of other hidden variables vary with $\delta$. The former case (\lq environmental quenching\rq) is known to be true, and is accounted for by fitting $f_q(\delta)$. The lack of density-dependence in the resulting correlation signal then indicates that the spatial correlation of hidden variables does not depend on $\delta$.

Finally, for case (iv), where $f_q=f_q(\mathrm{M_*},\delta)$, we have $\sigma_{\Delta}^2 = 0.2134\pm 0.0005$. Our \lq best prediction\rq\ of $q$ based on $(\mathrm{M_*},\delta)$ does not present a remarkable improvement over the \lq most ignorant\rq\ prediction from $\bar{q}$. This is an interesting point to note: although mass and environment are known to be important drivers of galaxy evolution, empirically accounting for their correlation with quenching does not account for much of the variance in $\Delta$. Concretely, this is due to the fact that while $f_q(\mathrm{M_*},\delta)$ does capture the tendency for galaxies to be quenched at high $\mathrm{M_*}$ or high $\delta$ (i.e. $f_q \sim 1$ in these regions), the majority of the galaxies in the sample, which have intermediate $\mathrm{M_*}$ and $\delta$, lie in the range where $f_q \sim 0.5$ (i.e. where the empirical model is \lq maximally ignorant\rq). There are two potential, mutually non-exclusive reasons for this. One possibility is that, as per the intuition developed with the toy model (Section \ref{subsec:toy_model}), there are other hidden variables which play a more important role to quenching. Another possible reason is that, even if quenching does depend \lq sharply\rq\ on $\mathrm{M_*}$ and $\delta$, quenched galaxies may ultimately migrate within variable space after the onset of quenching; this migration could simply be due to observational errors in the variables, or due to genuine evolution in these variables after quenching (e.g. further infall into an over-density, or residual star-formation within gas reservoirs). These effects can add scatter to an originally \lq sharp\rq\ dependence of quenching on $\mathrm{M_*}$ and $\delta$, which would then limit the extent to which they can empirically account for the variance in $\Delta$. Therefore, while these results indicate that $\mathrm{M_*}$ and $\delta$ do not provide strong empirical predictions of star-formation states, we caution against an over-interpretation of this without further analysis.

Measuring the correlation of $\Delta$ yields $\xi_{\Delta,0}=0.056\pm0.005$ (u.w.) and $0.060\pm0.009$ (d.w.), which correspond to $\gamma_\Delta=0.26\pm0.02$ (u.w.) and $0.28\pm0.04$ (d.w.). These ratios are smaller than those of (i), which suggests that, after taking into account the effects of $\mathrm{M_*}$ and $\delta$, the remaining spatially uncorrelated (\lq internal\rq) hidden variables play a slightly more important role in quenching than spatially correlated (\lq environmental\rq) ones. The spatial scale of this correlation is further reduced relative to (iii), to the point where there is no correlation beyond 3 Mpc. 

It is important to note that, since the statistical properties of $\Delta$ may vary with $\mathrm{M_*}$ and $\delta$, the results of the above analysis will in general depend on the distribution of the sample in variable space, and a sample which is selected differently may yield quantitatively different results. Therefore, it would be more meaningful to perform the analysis across different parts of the $(\mathrm{M_*},\delta)$ variable space. We do so by first splitting the sample into nine subsamples based on their position in variable space, as shown in Figure \ref{fig:sdss_subsets}. The boundaries of the selection are at $\mathrm{log}(\mathrm{M_*})=\left[9.5,10.20,10.55,11.25\right]$, and $\mathrm{log}(\delta)=\left[-4.0,-2.45,-1.65,0.00\right]$. The contours indicate the percentiles of the sample distribution at 20-percent intervals. The corresponding subsample autocorrelations of $\Delta=q-f_q(\mathrm{M_*},\delta)$ are shown in Figure \ref{fig:sdss_subcorrs}, where the positions of the panels match the positions of the divisions in Figure \ref{fig:sdss_subsets}, with $\mathrm{M_*}$ increasing from left to right, and with $\delta$ increasing from the bottom to the top. Given the discussion of case (iii), we only show the results for the unweighted signal; in any case, we have confirmed that the unweighted and the weighted signals do not differ significantly. To allow for meaningful comparison across different panels, the results are expressed in terms of $\gamma_\Delta(R)$ (i.e. as a fraction of the respective subsample variance of $\Delta$). Due to the larger uncertainties in these measurements, we do not attempt to deconvolve the measured signal. It should therefore be noted that, in the following, the strength of the correlation of $\Delta$ is likely to be systematically underestimated.

The first point to highlight from these results is a reiteration of a previous point, that accounting for the effects of mass and environment explains little of the variance of $\Delta$ in our sample. The majority of our sample lie in a region of $(\mathrm{M*},\delta)$-space where these two variables have little empirical predictive power, and thus the level of ignorance as quantified by the subsample variances is high.

We then consider what can be inferred about hidden variables in the low-mass regime. First, it is clear that the super-Mpc-scale correlation signal originates almost exclusively from low-mass galaxies (i.e. the leftmost three panels). Even in the sub-Mpc range, the full-sample correlation signal is mostly driven by these galaxies. That is, after accounting for the average effects  of $\mathrm{M*}$ and $\delta$, environmental effects (from environmental hidden variables) are still relatively important for the quenching of low-mass galaxies. This also indicates that $\delta$ alone, as measured by the $5^{th}$-nearest-neighbour method, is not sufficient to empirically describe the effects of \lq environment\rq\ on quenching.

On the other hand, for high-mass galaxies, we find no statistically significant correlation of $\Delta$ beyond 300 kpc, and relatively weak signals below that scale. That is, after accounting for the average effects of $\mathrm{M*}$ and $\delta$, the star-formation state of high-mass galaxies appear to have no further dependence on their environment, but rather depend primarily on their internal properties.

Overall, it can be said that the residual quenching (i.e. $\Delta$) of high-mass galaxies depend mostly on their internal properties, whereas the residual of low-mass galaxies have a strong dependence on their environment. Qualitatively, this observed trend of conformity (i.e. the magnitude and scale of $w_\Delta$) becoming stronger with decreasing mass resembles the measurement made by \citet{Kauffmann2013}, and may well be driven by issue (ii) discussed in Section \ref{subsec:meth_caution}. Namely, the few-Mpc-scale signal may simply reflect the quenching of satellites within very massive haloes. As discussed, the methodological modification which addresses issue (ii) necessarily requires the determination of the halo memberships of galaxies. Before we continue with a halo-based analysis of the observational data, we first measure the conformity signal in the mock. The motivation for doing so is twofold. First, the analysis in \cite{Sin2017} had indications that the SAM contained a conformity signal which was similar to that seen in the data. We would like to confirm this under our new framework. Second, as discussed in Section \ref{sec:concept}, this current work does not include a rigorous propagation of the errors associated with observational group identification. On the other hand, the group memberships of mock galaxies are relatively free from errors. Therefore, examining the mock and the data in parallel may aid in the interpretation of the results.

\subsection{Results from mock data}
\label{subsec:mock_res}

\begin{figure*}
\centering
\includegraphics[width=17.9cm]{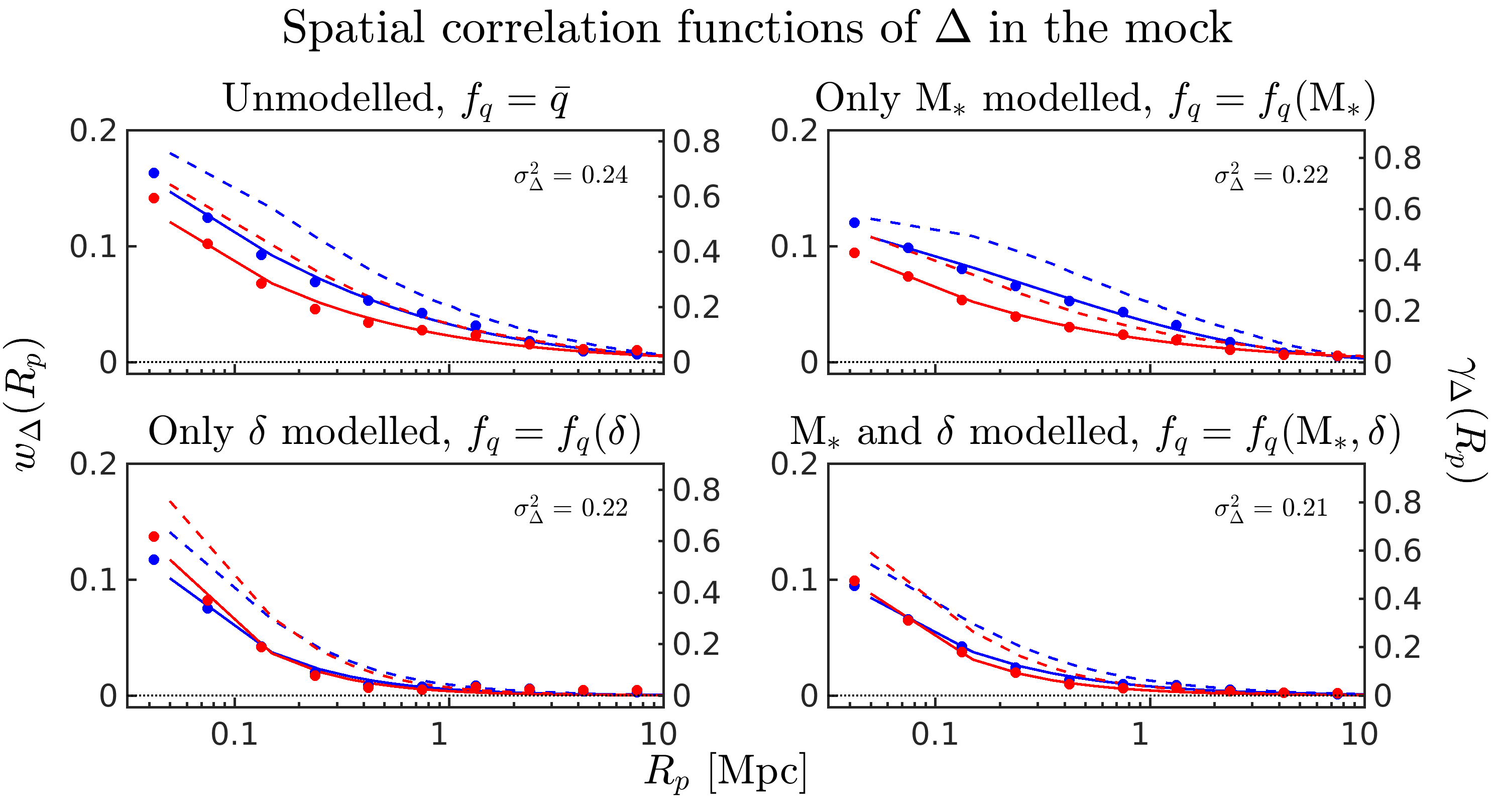}
\caption{Correlation functions of $\Delta$ measured from a mock catalogue constructed from the \citet{Henriques2015} SAM. The different panels, colour-coding, and the styling of the plots carry the same meaning as in Figure \ref{fig:sdss_fullcorr}, but note that the vertical axes span double the range.}
\label{fig:sam_fullcorr}
\end{figure*}

\begin{figure*}
\centering
\includegraphics[width=17.9cm]{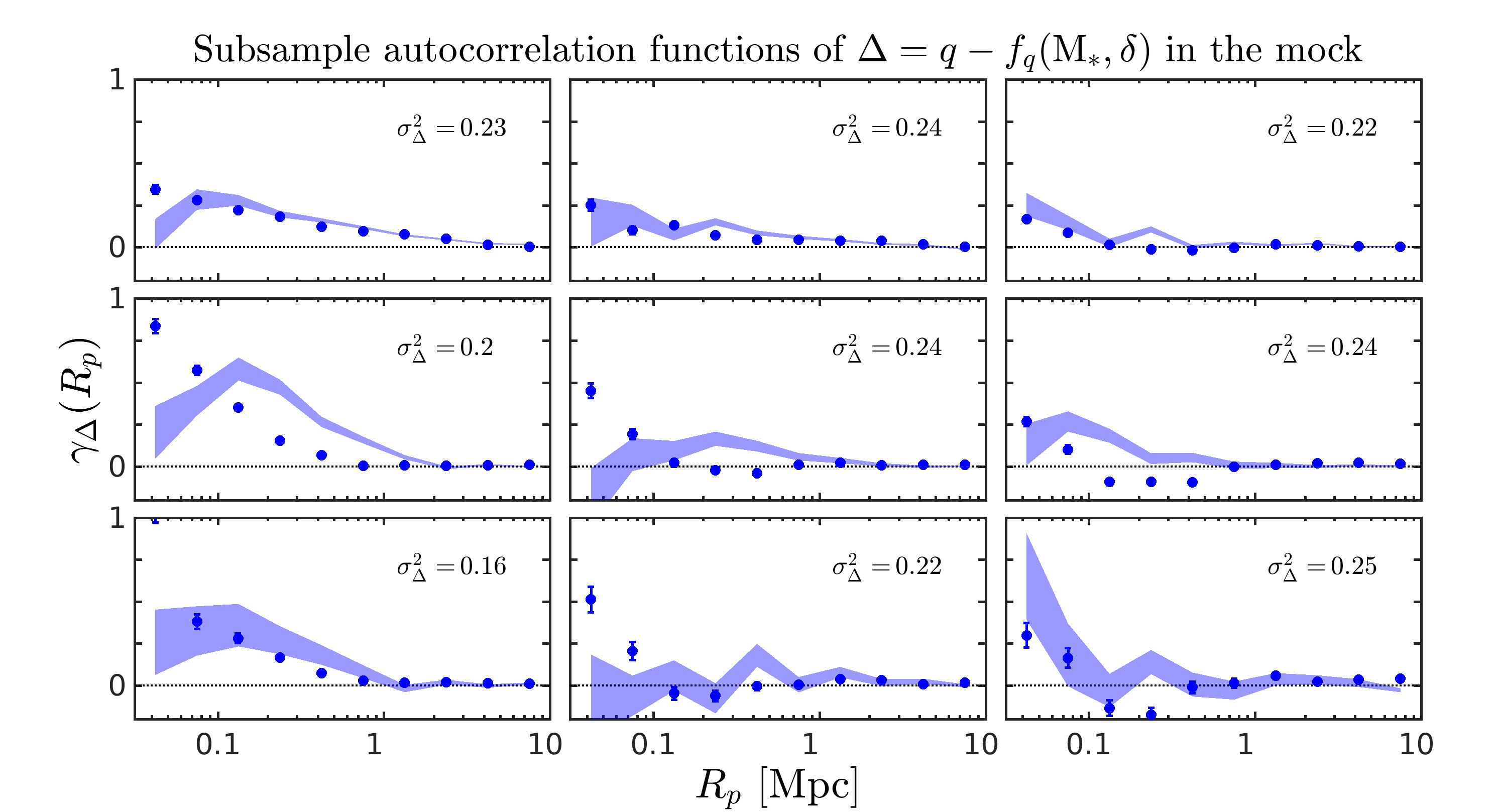}
\caption{The analogue of Figure \ref{fig:sdss_subcorrs}, measured from the mock catalogue. For comparison, the corresponding subsample autocorrelation functions measured from the SDSS data are plotted as shaded regions.}
\label{fig:sam_subcorrs}
\end{figure*}

We then apply an identical analysis to the mock dataset constructed from the H15 SAM.

The mock analogue of the full-sample correlation function (i.e. Figure \ref{fig:sdss_fullcorr}) is shown in Figure \ref{fig:sam_fullcorr}. The most striking difference to note is that, at small separations (below 100 kpc), the measured correlation in the mock is roughly double that in the SDSS data (note the factor of two difference in the range of the vertical axes). Beyond 100 kpc, the results from the mock are in fact consistent with those in the SDSS. The qualitative behaviour of $\Delta$ across the different panels is consistent with both the SDSS results, and with our physical understanding of these variables. However, as stated before, the quantitative results from these plots depend on the variable-space distribution of the underlying dataset; since the SDSS and the mock are distributed differently, a direct comparison between Figures \ref{fig:sdss_fullcorr} and \ref{fig:sam_fullcorr} is not meaningful. A better comparison would be that between the autocorrelations of subsamples selected in variable space.

The mock analogue of the autocorrelation of $\Delta$ within subsamples (i.e. Figure \ref{fig:sdss_subcorrs}) is shown in Figure \ref{fig:sam_subcorrs}, with their SDSS counterparts shown as shaded regions in every panel for comparison. There are some notable differences between the two. The first is that the sub-100 kpc correlation is stronger than that of the SDSS in the low- and intermediate-mass regime. However, given that the statistical uncertainties of the SDSS measurements are relatively high at this range, the significance of this difference is unclear. The second difference to note is that, between 0.1 and 1 Mpc, the signals from the mock tend to be suppressed relative to that from the SDSS. The most extreme examples of this are the bottom two panels of the rightmost column, where there is a significant anti-correlation signal. We have identified the cause of this anti-correlation to be the distinctly different prescriptions of quenching for centrals and satellites in the SAM; in the regions of variable space which correspond to these two bins, centrals tend to lie systematically above the $f_q(\mathrm{M_*},\delta)$ relation, and therefore have on average positive $\Delta$, while their satellites tend to lie below, and have on average negative $\Delta$. Moreover, central-satellite pairs originating from the same halo make up roughly half of all pairs at this separation in these two peculiar bins, and thus produce an anti-correlation on average. It is unclear whether this behaviour of centrals and satellites also explains the suppression of the signals relative to the SDSS results in the other $(\mathrm{M_*},\delta)$ bins. 

This example serves as an opportunity to reiterate an important idea in our framework: $\Delta$ and conformity are not intrinsic properties of the Universe. Rather, they reflect the incompleteness of our \lq best\rq\ attempt to describe quenching, as quantified by $f_q$. In this case, the anti-correlation indicates that our description based on $\mathrm{M_*}$ and $\delta$ fails to capture an important aspect of quenching in the SAM universe, namely that the quenching behaviour of centrals and satellites are distinctly different.

Other than these points, the results from the mock essentially agree with those from the SDSS; the overall behaviour of $\sigma_{\Delta}^2$ within each subsample is similar to that in the SDSS results, indicating that $\mathrm{M_*}$ and $\delta$ are also poor empirical predictors of the star-formation state of galaxies in the mock. On scales beyond 1 Mpc, the autocorrelation functions measured for subsamples in the mock also agree well with the SDSS results, and show once again that a large-scale correlation of the residual exists exclusively for low-mass galaxies.

For the mock, it is straightforward to verify that the majority of the low-mass galaxies are indeed satellites of very massive haloes, whose virial diameters are on average $\sim$2$\,\mathrm{Mpc}$, and in extreme cases out to $\sim$4$\,\mathrm{Mpc}$. This reinforces the suggestion in the previous subsection, that a conformity signal on any spatial scale is driven almost entirely by correlations within individual haloes, rather than by correlations across haloes.

\subsection{Interpretation of the correlation scale in the context of haloes}
\label{subsec:halo_interpretation}

\begin{figure*}
\centering
\includegraphics[width=17.9cm]{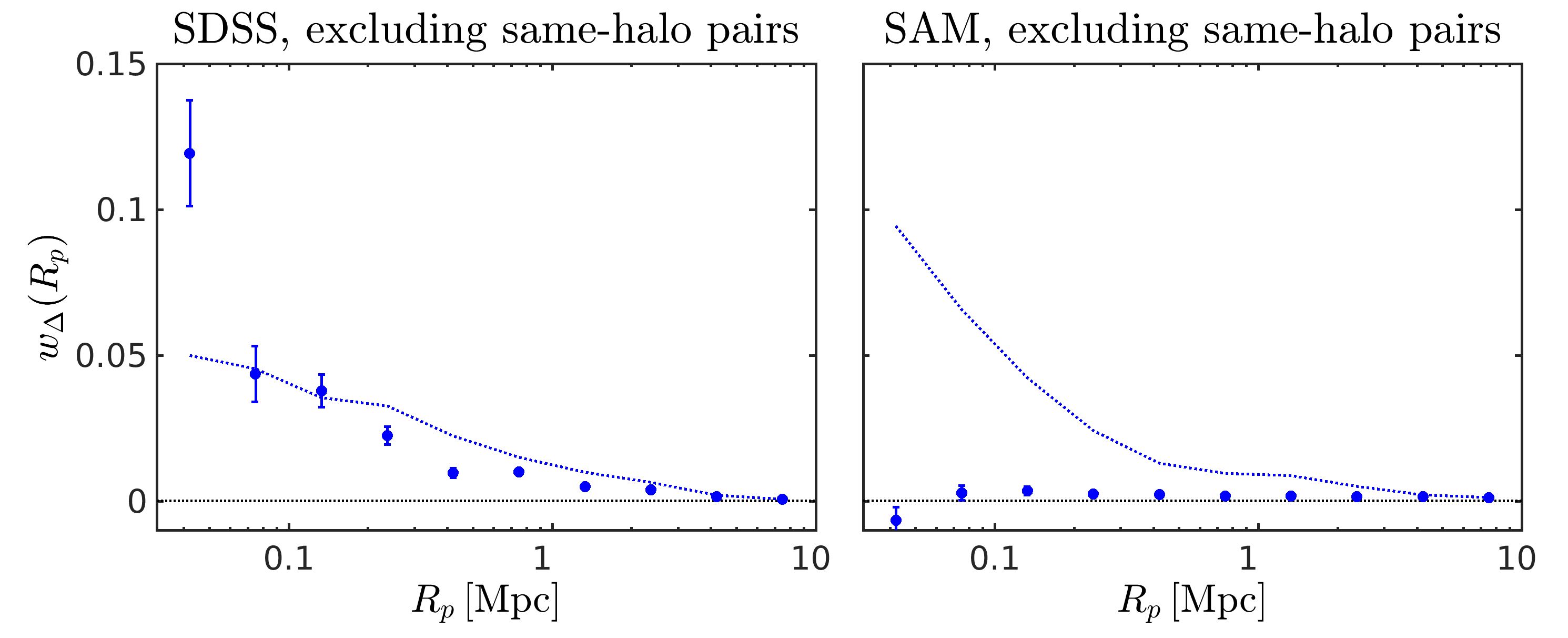}
\caption{An estimate of the \lq two-halo\rq\ component of the correlation functions shown in Figures \ref{fig:sdss_fullcorr} and \ref{fig:sam_fullcorr}, obtained by measuring the correlation function exclusively for pairs which are members of different groups. In practice, this is done by masking same-halo pairs from the existing full list of pairs; doing so masks of order 10 per cent of all pairs in the SDSS and the mock. The results from the SDSS sample is shown on the left, and those from the mock are on the right. For reference, the full-sample correlations from the respective datasets are plotted in dotted lines.}
\label{fig:crosshaloes}
\end{figure*}

Finally, we begin to address the role of haloes in this subsection. Given the group memberships and the list of pairs of a sample, it is straightforward to decompose the correlation function into its \lq one-halo\rq\ and \lq two-halo\rq\ components. To recap Section \ref{subsec:meth_caution}, the one-halo component of the full-sample signal can be measured by evaluating equation (\ref{eq:obs_corrfun}) exclusively for pairs in which both elements are members of the same halo. Similarly, the two-halo component can be measured by doing so exclusively for pairs whose elements are members of two different haloes.

We first consider the two-halo signal. By excluding same-halo pairs from the full list of pairs (i.e. leaving only different-halo pairs), one excludes of order 10 per cent of all pairs from the SDSS and the mock. The panels of Figure \ref{fig:crosshaloes} show the analogues of Figures \ref{fig:sdss_fullcorr} and \ref{fig:sam_fullcorr} under this modification. For comparison, the unweighted full-sample correlations for the respective datasets are plotted in dotted blue.

For the SDSS, the two-halo signal is somewhat weaker than the full-sample signal, and declines faster with increasing spatial scale, to the point where there is no two-halo correlation of $\Delta$ beyond 3 Mpc. However, it is also clear that on scales below 3 Mpc, a statistically significant two-halo correlation, which is not much weaker than the original full-sample signal, remains. That is to say that, although some of the Mpc-scale correlation in Figures \ref{fig:sdss_fullcorr} and \ref{fig:sdss_subcorrs} is indeed driven by \lq simple\rq\ halo quenching, there also appears to be a weak, genuine two-halo correlation of the residual. However, as the full-sample signal is a weighted average of the same-halo and different-halo signals, it follows that the same-halo signal is stronger than the two-halo signal at the $0.1-3\,\mathrm{Mpc}$ range. We emphasize that the significance of these statements ultimately depends on the accuracy of the group catalogue.

In the mock, there is no statistically significant two-halo correlation of $\Delta$. It is clear that, relative to $\mathrm{M_*}$ and $\delta$, the remaining spatially correlated (i.e. environmental) hidden variables must be properties directly related to individual haloes. That is, conformity in the SAM is driven purely by halo properties\footnote{Given that the prescriptions of quenching in the SAM are indeed halo-based, this is not at all surprising.}. 

These results suggest that any further investigation of the drivers of conformity should be centred on halo properties, rather than on $(\mathrm{M_*},\delta)$. However, Figure \ref{fig:crosshaloes} already indicates that, by directly using a group catalogue for a halo-based analysis of the SDSS data, we already measured highly discrepant results between the SDSS and the SAM. It is unclear whether these discrepancies are due to the statistics of quenching in the SAM being genuinely different from those in the Universe, or due to systematics introduced by errors in group identification. Therefore, we reserve a detailed analysis for an upcoming work, in which we will explore the origin of conformity in the context of haloes, both in the SDSS and in the SAM, while accounting for the systematic uncertainties in conformity measurements associated with the use of observational group catalogues.

\section{Discussion}
\label{sec:discussion}

An important development presented in this work is the conceptual framework for analysing the statistics of the star-formation state of galaxies, which is based on the statistic $\Delta$. While many other works have attempted to measure conformity in different settings, and with different methods, the framework that we have presented frames the problem in a general, meaningful, and self-consistent way. It does so through the consideration of the general idea of \lq unexplained\rq\ variance and correlation, which we attribute to the incompleteness of our understanding of galaxy evolution.

However, our quantitative understanding of the statistics of $\Delta$ is incomplete. In particular, the change in the values of $\sigma_{\Delta}^2$ and $\xi_{\Delta}$ as hidden variables are accounted for is not well-understood. Even for a simple, well-behaved toy model (Section \ref{subsec:toy_model}), one finds that the ratio of $\xi_{\Delta}$ and $\sigma_{\Delta}^2$ has an unintuitive form. It is unclear at this point whether tractable analytical relations exist for the general case (e.g. with many more variables which are likely to have complicated distributions, or where $q$ may have a complicated dependence on these variables), but this is nevertheless a subject that is being explored.

An important conclusion of this work pertains to the physical interpretation of the scale of conformity. Our results suggest that conformity for the most part is a halo-scale phenomenon. In the SDSS data, the correlation of $\Delta$ in the $0.1-3\,\mathrm{Mpc}$ range was found to be primarily due to same-halo pairs, although a weaker but significant two-halo correlation persists. In the mock, the spatial correlation of $\Delta$ can be attributed entirely to same-halo pairs. It is worth reiterating that these results emerged within a framework which made no assumptions that haloes, or central-satellite status, play a special role in quenching.

These results highlight some important outstanding issues regarding the scale of conformity. On one hand, they reaffirm the conclusion of \cite{Sin2017}, that as far as the data is concerned, there is only evidence for a relatively weak two-halo conformity, and one which is confined to the sub-$3\,\mathrm{Mpc}$ range. On the other hand, the existence of this two-halo correlation, albeit weak, is not expected given our current understanding of galaxy evolution, in the sense that it is completely absent from an identical analysis of the mock. The significance of this difference is unclear: as mentioned before, errors in the identification of groups in the group catalogue may introduce a spurious two-halo correlation. However, the differences between Figure \ref{fig:sdss_subcorrs} and \ref{fig:sam_subcorrs} suggest that there may also be genuine and significant differences between the Universe and the SAM in terms of how star-formation in galaxies are affected by halo properties.

Practically speaking, these results indicate that, while more difficult, a halo-based analysis is in fact necessary to further develop our understanding of quenching and conformity. In particular, this work raises at least two questions which we wish to explore. First, we wish to understand the role of grouping errors in group catalogues, and determine whether the two-halo signal in the SDSS sample is a result thereof, or if it reflects a genuine two-halo correlation. Second, we wish to understand the origin of the one-halo correlation in both the mock and the SDSS sample. We plan to do so by analysing the two datasets in parallel, and referring to the semi-analytical prescriptions in the mock for physical insight.

\section{Summary and conclusions}
\label{sec:summary}

In the first part of this work, we developed a conceptual framework and methodology with which one can meaningfully and quantitatively study the drivers of quenching. The framework is centred on the statistic $\Delta$, which is defined as the difference between the observed star-formation state of a galaxy, and our best prediction of its state based on our current empirical understanding of quenching. In particular, this work uses the average quenching effects of stellar mass $\mathrm{M_*}$ and local galaxy number density $\delta$ to approximately express our current understanding. We think of $\Delta$ as a residual which reflects the effects of drivers of quenching which are not captured by $\mathrm{M_*}$ and $\delta$, or so-called \lq hidden variables\rq.

Through a toy model, we explored how the statistical properties of $\Delta$ reflect the quenching effects of internal and external hidden variables of a sample of galaxies. In particular, the variance of $\Delta$ reflects the effects of hidden variables in general. On the other hand, the spatial correlation function of $\Delta$ exclusively reflects the effects of spatially correlated (i.e. environmental) hidden variables; we further argue that the correlation function is a more general and meaningful quantification of conformity. 

When measuring conformity in this way, there are methodological issues which, if not understood, could lead one to misinterpret the measurement. These issues of projection, density-weighting bias, and the difficulty in interpreting the scale of a measured signal are explained and addressed.

We then applied this method of analysis to a sample of local galaxies selected from the Sloan Digital Sky Survey, and find that after accounting for the average quenching effects of $\mathrm{M_*}$ and $\delta$, there remains a significant residual $\Delta$ which has relatively high variance, and which is correlated out to separations of roughly 3 Mpc. By further analysing the properties of $\Delta$ within subsamples selected in $\mathrm{M_*}$ and $\delta$, we find that environmental hidden variables remain important for driving the residual quenching of low-mass galaxies, while the residual quenching of high-mass galaxies is driven mostly by internal properties. These results, along with a parallel analysis of a similarly-selected sample from the Henriques et al. (2015) mock, strongly suggest that it is necessary to consider halo mass, and other halo-related properties, as candidates for hidden variables. 

A preliminary halo-based analysis indicates that, in the SDSS data, there is indeed a genuine two-halo correlation of $\Delta$ out to at most 3 Mpc, albeit one which is weak relative to the one-halo correlation at that range of separation. The statistical significance of this signal is unclear, and requires an in-depth analysis of the systematics introduced by group catalogues. On the other hand, within the mock, much of the variance and all of the correlation of $\Delta$ can be attributed to the physics associated with individual haloes. However, it is also unclear what the hidden variable driving this behaviour is. These open issues will be addressed in an upcoming work.

\section*{Acknowledgements}

This work has been supported by the Swiss National Science Foundation. BMBH (ORCID 0000-0002-1392-489X) acknowledges support from an ETH Zwicky Prize Fellowship. 

\bibliographystyle{mn2e} \bibliography{SLH2019}
\label{lastpage}

\end{document}